\DocumentMetadata{}
\documentclass[acmlarge]{acmart}
\pdfoutput=1 

\usepackage{longtable}
\usepackage{pgfplots}
\usepackage{color}
\usepackage{xcolor}
\usepackage{listings}
\usepackage{xparse}
\usepackage{makecell}
\usepackage{hyperref}
\pgfplotsset{width=10cm,compat=1.9}
\usepackage{multirow}
\usepackage{subcaption}
\usepackage{listings}

\definecolor{codegreen}{rgb}{0,0.6,0}
\definecolor{codegray}{rgb}{0.5,0.5,0.5}
\definecolor{codepurple}{rgb}{0.58,0,0.82}
\definecolor{backcolour}{rgb}{0.95,0.95,0.92}

\lstdefinestyle{mystyle}{
    backgroundcolor=\color{backcolour},   
    commentstyle=\color{codegreen},
    keywordstyle=\color{magenta},
    numberstyle=\tiny\color{codegray},
    stringstyle=\color{codepurple},
    basicstyle=\ttfamily\footnotesize,
    breakatwhitespace=false,         
    breaklines=true,                 
    captionpos=b,                    
    keepspaces=true,                 
    numbers=left,                    
    numbersep=5pt,                  
    showspaces=false,                
    showstringspaces=false,
    showtabs=false,                  
    tabsize=2
}

\usepackage{hhline}
\usepackage{array}
\newcolumntype{L}[1]{>{\raggedright\let\newline\\\arraybackslash\hspace{0pt}}m{#1}}
\newcolumntype{C}[1]{>{\centering\let\newline\\\arraybackslash\hspace{0pt}}m{#1}}
\newcolumntype{R}[1]{>{\raggedleft\let\newline\\\arraybackslash\hspace{0pt}}m{#1}}

\AtBeginDocument{%
  \providecommand\BibTeX{{%
    \normalfont B\kern-0.5em{\scshape i\kern-0.25em b}\kern-0.8em\TeX}}}

\acmJournal{TOSEM}
\acmYear{2024} \acmVolume{33} \acmNumber{0} \acmArticle{0} \acmMonth{0} 
\renewcommand{\shortauthors}{Amareen \em{et} al.}

\begin{document}

%% allowing the author to define a "short title" to be used in page headers.f
\title{GraphQL Adoption and Challenges: Community-Driven Insights from StackOverflow Discussions}

\author{Saleh Amareen}
\email{saleh.amareen@wayne.edu}
\affiliation{
  \institution{Wayne State University}
  \streetaddress{42 W Warren Ave}
  \city{Detroit}
  \state{Michigan}
  \country{USA}
  \postcode{48202}
}

\author{Obed Soto Dector}
\email{obed.sotodector@dominos.com}
\affiliation{
  \institution{Domino's}
  \streetaddress{30 Frank Lloyd Wright Dr}
  \city{Ann Arbor}
  \state{Michigan}
  \country{USA}
  \postcode{48106}
}

\author{Ali Dado}
\email{ali.dado@dominos.com}
\affiliation{
  \institution{Domino's}
  \streetaddress{30 Frank Lloyd Wright Dr}
  \city{Ann Arbor}
  \state{Michigan}
  \country{USA}
  \postcode{48106}
}

\author{Amiangshu Bosu}
\email{amiangshu.bosu@wayne.edu}
\orcid{1234-5678-9012}
\affiliation{
  \institution{Wayne State University}
  \streetaddress{42 W Warren Ave}
  \city{Detroit}
  \state{Michigan}
  \country{USA}
  \postcode{48202}
}

\newcommand{\bnote}[1]{\textcolor{red}{[Bosu:#1]}}
\newcommand{\anote}[1]{\textcolor{green}{[Saleh:#1]}}

\newenvironment{boxedtext}
    {
    
    \begin{center}

    \renewcommand{\arraystretch}{1.15}
    \begin{tabular}{|p{0.96\linewidth}|}
    \hline
    }
    { 
    \\ \hline
    \end{tabular} 
    \renewcommand{\arraystretch}{1}
    \end{center}
       }

\renewcommand{\shortauthors}{Amareen \em{et} al.}
\definecolor{MidnightBlue}{rgb}{0.1, 0.1, 0.44}
\newcommand{\post}[1]{\textcolor{MidnightBlue}{\textit{#1}}}
\newcommand{\code}[1]{{\tt #1}}

\begin{abstract}
%GraphQL is a query language and web application programming interface (API) for client-server architecture. Its advantages include type-safe queries, which allow clients to retrieve the data they require precisely in a single request. 
As organizations adopt GraphQL for API implementations, it is imperative to understand its challenges and the software community's interests. To achieve this goal, we conducted a five-step mixed-method empirical analysis of 45K StackOverflow questions and answers on GraphQL. In the first step, we derive a reference architecture for the GraphQL ecosystem with five key layers. Second, we used topic modeling based on Latent Dirichlet Allocation (LDA) to automatically identify 14 topics and 47 subtopics. Third, we mapped discussion topics to architecture layers. Fourth, we manually investigate questions on each topic and subtopics to provide additional insight to the GraphQL stakeholders. Finally, we study topic difficulty, popularity, trends, and tradeoffs to provide insights into evolving community interests and challenges. Our results indicate that Client and Server are the top two architectural layers attracting discussion on SO. While earlier discussions on SO focused on building third-party applications consuming GraphQL APIs (i.e., API Integration) released by large organizations, recent trends suggest more organizations implementing APIs using GraphQL servers. Due to difficulty and lack of well-defined solutions, security remains a difficult and low-interest area. However, such a practice can lead to vulnerable APIs.
\end{abstract}



\maketitle

\section{Introduction} \label{sec:introduction}

% What is GraphQL
In 2012, Facebook developed GraphQL, a query language and an application programming interface (API) for web-based client-server architectures. Facebook had internally utilized GraphQL and continued evolving its formal specification before finally open-sourcing it and releasing its specification and implementation to the public in 2015. Facebook then transferred GraphQL to the non-profit GraphQL Foundation in late 2018. 
GraphQL has gained a rapid adoption momentum in recent years, especially because it provides a technology paradigm that is either alternative or additional to existing web protocols such as Representational State Transfer (REST) and Simple Object Access Protocol (SOAP)~\cite{controlledexperiment}. While the latter protocols dominated the industry for the past decade, GraphQL has recently been increasingly adopted by major software companies such as Shopify, Amazon, Pinterest, Netflix, and Twitter~\cite{graphqlformaldef2, replacerest,controlledexperiment}. A recent Gartner report predicts that at least 50\% of enterprises will be utilizing GraphQL in production by 2025, which stood at 10\% in 2021~\cite{ibm_gartner_blog_post}. 
%One of the earliest adopters of GraphQL is Github, who started their GraphQL migration as early as 2016 in conjunction with their existing REST APIs~\cite{replacerest}.

% Key properties of GraphQL 
One of the most prevalent properties of GraphQL is that clients can form type-safe GraphQL-like queries or mutations that declaratively and precisely specify the data they require in a single call. This feature allows clients to avoid issues commonly referred to as over-fetching and under-fetching that developers encounter while using legacy protocols such as REST. Over-fetching occurs when responses of REST API calls include more data than clients require. On the contrary, under-fetching occurs when existing APIs provide less data than clients require, forcing clients to perform multiple calls to aggregate all necessary data. Over and under-fetching of data represents a sub-optimal utilization of resources.

Although many organizations have increasingly adopted GraphQL, developers encounter numerous challenges due to technical complexities and inadequate resources. One of the key support mechanisms for GraphQL developers is StackOverflow (SO), which is a well-known knowledge-based website and part of the StackExchange (SE) network~\cite{stackexchangesites}. SO moderates many posts that software professionals and enthusiasts discuss in the form of questions and answers (Q\&As) regarding various programming topics. A systematic analysis of GraphQL-related questions will help the GraphQL community, including developers, told designers, and researchers, identify areas requiring support and effort.  

\begin{figure*}[t!]
    \centering
    \begin{subfigure}[t]{0.48\textwidth}
        \centering
         \includegraphics[width=\linewidth]{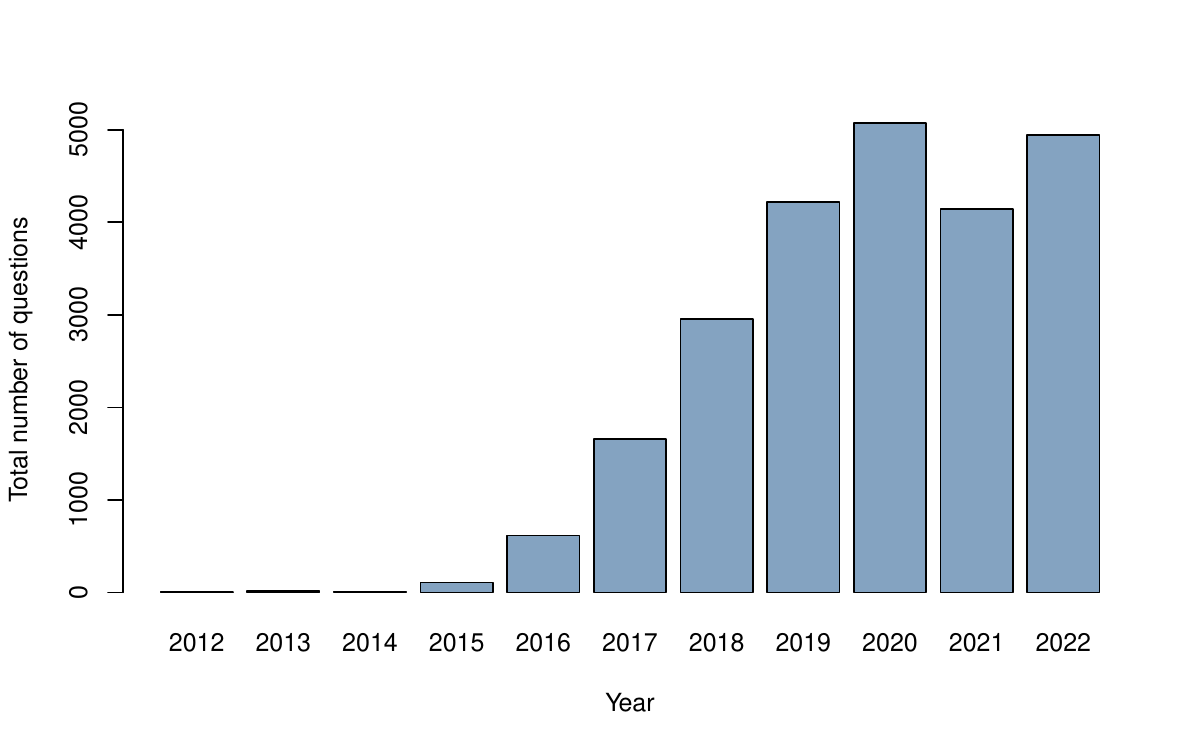}
  \caption{Total number of GraphQL-related questions on the StackOverflow website per year, a higher number indicates growing popularity.}
  \label{fig:total_graphql_questions_per_year}  
    \end{subfigure}%
    ~ 
    \begin{subfigure}[t]{0.48\textwidth}
        \centering
        \includegraphics[width=\linewidth]{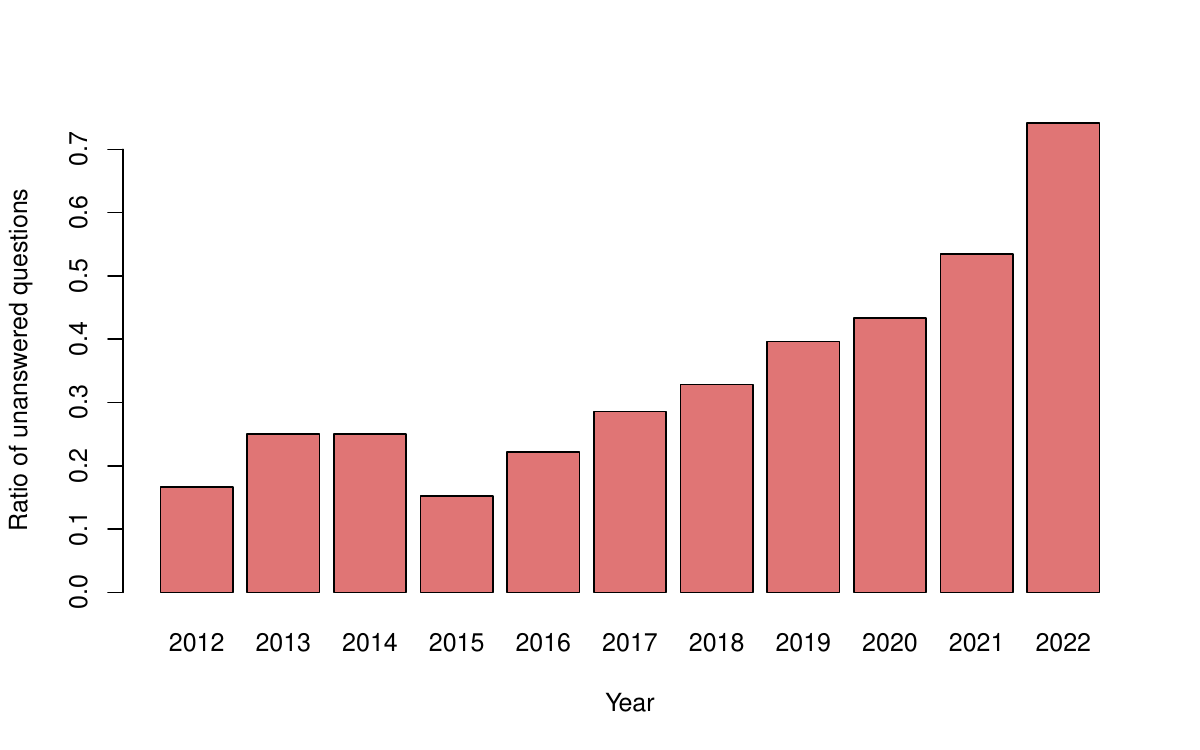}
  \caption{Ratio of GraphQL-related questions without accepted answers on the StackOverflow per year. A higher number indicates growing difficulty.}
   \label{fig:ratio_unanswered_graphql_questions_per_year}

    \end{subfigure}
    \caption{Questions and answers on StackOverflow over with \#graphql tag over the years. }
\end{figure*}

\vspace{4pt}
\noindent \textbf{Motivating preliminary analysis:}
Before committing to a full-scale investigation, we conducted a preliminary study with an SO data dump from December 2022 with the posts with the most prevalent GraphQL tag (i.e., \#graphql). Figure~\ref{fig:total_graphql_questions_per_year} shows that software developers' total number of questions on SO increased annually between 2012 and 2019. Since the number of questions has steadily fluctuated between 4K and 5K per year, this finding reinforces previous research work that GraphQL continues to gain momentum and indicates that developers continue to encounter challenges that require further assistance from the GraphQL community.
Our second finding, represented in Figure~\ref{fig:ratio_unanswered_graphql_questions_per_year}, shows that the ratio of unanswered questions (i.e., questions with no accepted answers) has been steadily increasing from 25\% since its public release in 2015 to 74\% in 2022. Such a finding indicates the increasing difficulty of the GraphQL domain in the GraphQL community. Sri \textit{et al.}~\cite{replacerest} confirms such difficulty of GraphQL in their recent study, which surveyed 38 Github employees who expressed that GraphQL was challenging to learn, time-consuming with limited knowledge-base resources, especially when compared to mature protocols such as REST. 
%Additionally, despite the abundance of existing research work that studies GraphQL~\cite{replacerest, controlledexperiment, migrating_longterm, morphgraphql, migratingtographql_apracticalassessment}, to the best of our knowledge, no prior work has yet explored the difficulty of GraphQL nor has any work categorized the community interest and evolution of various discussions topics ever since its first release to the public in 2015. 
Moreover, the GraphQL domain is vast with programming-language-specific implementations, libraries, frameworks, tools, IDEs, APIs, platforms, and documentation~\cite{graphqltools}. It is unknown whether such breadth may leave GraphQL developers, software organizations, researchers, and educators prone to several challenges, especially for newcomers. Therefore, it is incumbent to aid the GraphQL community in understanding the difficulties, popularity, and trends of the various GraphQL topics because such understanding will help its stakeholders make informed decisions, identify opportune research areas, and guide where they should focus their time and efforts.

% Research Questions
\vspace{4pt}
\noindent \textbf{Objective and approach:}
This paper aims to \textit{help the GraphQL community of practitioners, researchers, and educators understand the architecture, interest, challenges, and trends of topics within the GraphQL ecosystem} based on software developer discussions on StackOverflow~\cite{stackoverflow}. 
To achieve this goal, we conducted a five-step mixed-method empirical analysis of 45K StackOverflow questions and answers on GraphQL. In the first step, we derive a reference architecture for the GraphQL ecosystem with five key layers based on two well-known GraphQL implementations. Second, we used topic modeling based on Latent Dirichlet Allocation (LDA) to automatically identify 14 topics and 47 subtopics.
LDA has been the de-facto standard in the software engineering (SE) discipline to analyze SO data~\cite{iot, bigdata, blockchain}. Numerous research works have conducted LDA-based empirical studies to analyze software developer discussions obtained from the stack exchange websites aiming to investigate domains such as big data~\cite{bigdata}, security~\cite{security}, blockchain~\cite{blockchain}, mobile development~\cite{mobile}, concurrency~\cite{concurrency}, code refactoring~\cite{refactor}, Open-Source licensing~\cite{opensource} and internet of things (IoT)~\cite{iot}. However, no prior research has analyzed developer discussions in the GraphQL domain.
Third, we mapped discussion topics to architecture layers. Fourth, we manually investigate questions on each topic and subtopics to provide additional insight to the GraphQL stakeholders. Finally, we study topic difficulty, popularity, trends, and tradeoffs to provide insights into evolving community interests and challenges. 

\vspace{4pt}
\noindent \textbf{Contribution:}
The primary contributions of this study are as follows.
\begin{enumerate}
    \item An empirical analysis of 45,554 GraphQL-related questions and answers on StackOverflow.
    \item A reference architecture for the GraphQL ecosystem using two popular GraphQL implementations. The architecture includes GraphQL client and server and describes their respective layers and components.
    \item A quantitative empirical study on the obtained GraphQL posts using topic modeling.  
    \item A mapping the GraphQL topics to their appropriate architecture layers and components of our derived reference architecture of the GraphQL ecosystem.
    \item A new approach to build multi-level topic hierarchy. 
    \item An empirical characterization of the discovered GraphQL topics, their difficulty, popularity, tradeoffs, and evolution.
    \item A set of implications and recommendations for the various GraphQL stakeholders.
    \item To facilitate replication and/or extension of this work by future researchers, we make our dataset publicly available~\cite{dataset_replication}\footnote{\url{https://github.com/WSU-SEAL/graphql-topic-modeling}}.
\end{enumerate}

\vspace{4pt}
\noindent \textbf{Organization:}
{The remainder of this paper is organized as follows}. Section~\ref{sec:related_work} provides a brief introduction to GraphQL and discusses related work. 
Section~\ref{sec:research-questions} introduces the five research questions of this study. 
We elaborate on our research methodology in Section~\ref{sec:methodology}. 
Section~\ref{sec:results} details the results of this study. Section~\ref{sec:discussion} discusses our results and suggests recommendations for the various GraphQL stakeholders. We present threats that impact the validity of our work in Section~\ref{sec:threats} before concluding in Section~\ref{sec:conclusion}.

\section{Background} \label{sec:related_work}

The following subsections provide a brief overview of GraphQL core concepts, research work on GraphQL, and software engineering studies on topic modeling.

\subsection{GraphQL}
 
To comprehend GraphQL,  we need to understand the following four core concepts.

\begin{lstlisting}[language=Python, caption=A Simple GraphQL Schema]
# Definition of a type called 'User'
type User {
  id: ID!       # A non-nullable unique identifier for the user
  name: String! # A non-nullable name of the user
  email: String # The email of the user, can be null
}

# Definition of a query type to fetch data
type Query {
  # A query to get a user by their ID
  user(id: ID!): User
}

\end{lstlisting}

\noindent \emph{1. GraphQL schema}: A GraphQL schema is a blueprint that defines the data structure in a GraphQL API. It outlines the types of data that can be queried, the relationships between those types, and the operations (queries, mutations, and subscriptions) that can be performed. Listing~ 1 shows an example GraphQL schema.

\vspace{3pt}
\noindent \emph{2. GraphQL Query language (GQL)}: GQL is used to interact with a GraphQL API. It allows clients to request exactly the data they need, and nothing more, in a structured and predictable way. This language is used to construct queries, mutations, and subscriptions to interact with the API.  Queries let clients retrieve data from the server, mutations support data modification, and subscriptions enable clients to subscribe to real-time events from a server. Listing 2 shows an example GraphQL query for the above schema.

\begin{lstlisting}[language=Python, caption=GraphQL query]
# Query to fetch email address of the user with a specific id
{
  user(id: "1") {
    email  
  }
}



\end{lstlisting}

\vspace{3pt}
 \noindent \emph{3. GraphQL Resolvers}: GraphQL resolvers are functions on the server side to fulfill the runtime execution of client requests to fetch, modify, or subscribe to resources. Resolvers may integrate with various data sources such as databases, REST services, or other APIs.

\vspace{3pt}
 \noindent \emph{4. GraphQL Lifecycle}:
The lifecycle of a GraphQL request begins when a client sends a query, mutation, or subscription to a GraphQL server endpoint. The server then validates the request against its schema. If the request is invalid, the server returns an error and does not execute the operation. If the request is valid, the server maps it and its nested fields to their corresponding resolvers for execution. Each resolver processes its part of the request and passes it down to the next level of resolvers until all requested fields have been handled. Finally, the resolver returns the data to the GraphQL endpoint, which sends the response back to the client, typically in JSON format.

\subsection{GraphQL Studies}
This section focuses on prior academic research on the GraphQL language specification, comparative studies that focus on GraphQL vs. REST and migration paths, testing, and solutions proposed to address the challenges software developers face while learning, using, or adopting GraphQL. Harting and Perez conducted one of the earliest works to provide an initial formal definition of the GraphQL language~\cite{graphqlformaldef1}. The authors later followed up with a second work~\cite{graphqlformaldef2} where they reported that the problem of evaluating the complexity of GraphQL queries is nondeterministic logarithmic space (NL). While Harting and Perez utilized query normalization to work with simpler semantics and were able to produce their GraphQL query complexity results, they did not provide a normalization procedure. To address that, Tomas \textit{et} al. proposed a mechanized formalization of GraphQL, which they implemented in CoQ proof assistant and named it GraphCoQL. Their work proposed and formally proved a normalization algorithm for Harting and Perez's query transformations and interpretations and proved them correct~\cite{mechanized}. Gleison \textit{et} al. conducted a literature review paper to characterize the advantages and disadvantages of GraphQL in the existing literature~\cite{migratingtographql_apracticalassessment}. In the same work, the authors employed the literature review results to assess the practicality of migrating from REST to GraphQL on seven systems. They showed GraphQL could significantly reduce the number of API calls and the number of fields in the returned JSON document and, therefore, reduce the overall returned payload size in bytes. Sri \textit{et al.} conducted a qualitative and quantitative empirical study to explore the replaceability of REST with GraphQL in the future~\cite{replacerest}. They surveyed 38 Github employees to understand the advantages and disadvantages of both protocols and used metrics such as efficiency and application feasibility to provide proper recommendations. They reported several key findings regarding the challenges faced by software developers while adopting GraphQL, such as its difficulty in learning, time consumption, and limited resources. They reported both protocols had their benefits and weaknesses.

On the other hand, a controlled experiment by  Brito and Valente with 22 students found client-side application integration with a GraphQL API easier than with REST~\cite{controlledexperiment}.  They compared the implementation of remote service queries using REST and GraphQL. They found that GraphQL required less effort, with a median implementation time of 6 minutes compared to 9 minutes for REST. Additionally, implementing REST queries became more challenging with complex endpoints and multiple parameters. Participants without experience found GraphQL easier to implement and understand~\cite{controlledexperiment}. 
Wittern \textit{et} al.'s empirical investigation of over 8K GraphQL schemas mined from GitHub suggests that most APIs have security issues that can be exploited using complex queries~\cite{wittern-2019}. 
 For GraphQL testing, Asma \textit{et al.} recently proposed a white-box testing framework to automate GraphQL testing employing code generation that is based on evolutionary search~\cite{evolutionarytesting}. The work aimed to maximize code coverage and improved it by up to +55.53\%. While other GraphQL automation testing research works had previously existed (e.g., automation based on deviation testing ~\cite{deviationtesting} and property-based testing~\cite{propertybasedtesting}), they had limitations in terms of their inability to analyze byte-code. Patrick \textit{et al.} proposed a non-intrusive and technology-independent GraphQL federation~\cite{graphqlfederation} to address the shortcomings of the most popular Apollo federation implementation~\cite{apollofederation}. Their work employed domain-specific language (DSL) traceability links to describe and consolidate multiple and possibly conflicting GraphQL schema. The authors provided a proof-of-concept implementation and a qualitative comparison of their features compared to the Apollo Federation.   Another recent direction focuses on assessing execution strategies and costs of GraphQL queries~\cite{cha-FSE-2020,mavroudeas-ASE-2021,roksela2020evaluating}.

\subsection{Topic Modeling in Software Engineering}
Many prior studies have used LDA-based topic modeling to analyze questions and answers on StackExchange and StackOverflow~\cite{bigdata, blockchain, iot, mobile, security, opensource}. Barua \textit{et.} al. conducted the first study in this direction using a StackOverflow data dump from September 2010~\cite{barua2014developers}. 
However, with the growing popularity of StackOverflow and its sister sites, subsequent studies have focused on specific areas by filtering questions using \code{\#tags}.  
For example, Christoffer and Emad analyzed 13 million mobile developer discussions on StackOverflow regarding iOS, Android, and Windows phone platforms to investigate difficult and popular mobile development topics~\cite{mobile}.
Bajaj \textit{et al.}~\cite{bajaj} presented a study to understand challenges and misconceptions web developers commonly encounter. Mehdi et al. ~\cite{bigdata} developed a topic hierarchy to understand big data-related discussions~\cite{bigdata}. 
Panichella \textit{et al.}~\cite{panichella} proposed a variation of LDA that utilized a genetic algorithm (LDA-GA) aiming to cluster topics optimally.
Yang \textit{et al.} studied security-related StackOverflow posts utilizing LDA-GA~\cite{security}. 
Other studies in this area include software refactoring~\cite{refactor}, IoT~\cite{iot}, web servers~\cite{web_server_ref_arch}, web browsers~\cite{web_browser_ref_arch}, blockchain~\cite{blockchain}, Open Source licenses~\cite{opensource}, AutoML toolkits~\cite{al2024far}, DevOps challenges~\cite{tanzil2023mixed}, 
low code software development~\cite{alamin2023developer}, platforms~\cite{alamin2023developer}, and
new programming languages~\cite{chakraborty2021developers}.

\vspace{3pt}
\noindent \textbf{Novelty:} This work differs from prior works in three key aspects. First, we focus on GraphQL, an emerging technology expected to see rapid adoption among enterprises~\cite{ibm_gartner_blog_post}.  Hence, understanding the challenges and needs of the GraphQL community is necessary to assist those adopters. To our knowledge, this analysis is currently missing. Second, we develop a reference architecture of the GraphQL ecosystem and map discussion topics on SO to these layers.  Finally, we utilize a new approach to build a topic hierarchy model to identify discussion sub-topics and present a three-level hierarchy of discussion areas. 

\section{Research questions}
\label{sec:research-questions}
This section introduces the five research questions driving this study with a brief motivation for each.

\vspace{2pt}
\noindent \textbf{(RQ1) How can we define a referenced architecture of the current GraphQL ecosystem?}

\noindent \underline{Motivation:} The wide spectrum of the GraphQL ecosystem contains an immense amount of documentation. Moreover, the abundance of technology stacks, programming-language implementations, frameworks, libraries, development tools, domain terms, and documentation complicates software developers' learning process. For example, the following SO post shows how current documentation confuses a software developer: \post{"...I came across various terms associated with React - Flux, Redux, Relay, GraphQL...and am confused as to which architecture to invest time and effort in learning and implementing..."} [\hyperlink{https://stackoverflow.com/questions/36772960/confused-with-react-patterns-architectures}{$Q_{36772960}$}]. Another example is this SO post title: \post{"is there an apollo graphl site official (apollo client , apollo server), what is the correct name?"} [\hyperlink{https://stackoverflow.com/questions/45647122/is-there-an-apollo-graphl-site-official-apollo-client-apollo-server-what}{$Q_{45647122}$}].
We aim to define a reference architecture for the current GraphQL ecosystem to help improve the learning curve, especially for newcomers to GraphQL. 

\vspace{2pt}
\noindent \textbf{(RQ2) What do GraphQL developers discuss on StackOverflow?}

\noindent \underline{Motivation:}  
Identification of discussion topics can help GraphQL stakeholders, such as its namesake foundation or its tool builders, understand common challenges practitioners encounter. These insights can assist them in pinpointing recurring problems needing improved design, documentation, FAQs, or tutorials. Moreover, these insights can help identify features or improvements most relevant to users and guide the development roadmap.

\vspace{2pt}
\noindent \textbf{(RQ3) How can we organize GraphQL discussions on StackOverflow into a hierarchy of topics?}

\noindent \underline{Motivation:} Each primary discussion area may have several sub-areas. For example, discussions on a particular tool may be divided into deployment, usage challenges, or integration issues. We aim to provide a finer-grained analysis with a multi-level topic hierarchy to better understand each area.

\vspace{2pt}
\noindent \textbf{(RQ4) How did GraphQL discussion topics on StackOverflow evolve?}

\noindent \underline{Motivation:} 
By analyzing discussion trends over time, we can observe how the challenges and interests around GraphQL evolve. This insight will allow the community to focus on current issues rather than outdated or less relevant topics. Moreover, the emergence of new discussions can highlight innovative uses of GraphQL, new patterns, or integrations that might not have been considered before.

\vspace{2pt}
\noindent \textbf{(RQ5) What are the difficulty and popularity of various GraphQL topics?}

\noindent \underline{Motivation:} This analysis will identify areas where the GraphQL community has better support and where it lags.  These insights can guide framework builders to prioritize new features or improvements based on what users are increasingly asking about or struggling with.

\section{Research Methodology} 
\label{sec:methodology}
The following subsections detail our research methodology for deriving current GraphQL ecosystem reference architecture, identifying discussion topics from the SO data dump, assigning names to topics, and mapping each topic to architecture layers.

\subsection{Deriving Reference Architecture of current GraphQL Ecosystem}
\label{methodology:derive_ref_architecture}
GraphQL ecosystem documentation comes from various sources. The GraphQL language specification~\cite{graphql_spec} is the primary documentation source for GraphQL building blocks that various GraphQL implementations comply with, whereas Apollo~\cite{apollo} represents the reference GraphQL implementation for both GraphQL client and server. Additionally, the GraphQL specification is (by design) agnostic of the underlying programming language. Therefore, libraries, frameworks, and various implementations across programming languages (e.g., Java, .Net, Python, and Javascript) have their documentation. Moreover, since GraphQL emerged, software organizations have offered GraphQL APIs in parallel or as alternatives to legacy RESTful APIs. Github, GitLab, Yelp, and Shopify are a few examples of APIs that do so. Lastly, there are currently many third-party vendor platforms that aim to provide a SaaS GraphQL experience, such as Amazon AppSync~\cite{appsync}, Hasura~\cite{hasura}, Prisma~\cite{prisma} and Gatsby~\cite{gatsby}. Our approach is motivated by reference architecture definition approaches from prior works~\cite{web_browser_ref_arch, web_server_ref_arch, blockchain} and is based on the following two steps.

\begin{enumerate}
    \item \textit{Derive a conceptual Architecture}: First, we derive a conceptual architecture utilizing the available GraphQL documentation as well as the domain knowledge of the author. The list of authors includes three members, each with more than five years of professional experience developing and maintaining large-scale enterprise GraphQL deployments. Subsequently, we reverse engineer the source code of two GraphQL implementations to refine our conceptual architecture and produce two concrete GraphQL architectures. The source-code implementations we choose are, namely, Apollo GraphQL~\cite{apollo} and the Domain Graph Service (DGS) framework by Netflix~\cite{dgs}. We selected these two since Apollo is the reference GraphQL implementation, whereas DGS is a Java implementation popular for enterprise microservices. Moreover, both Apollo and DGS provide client and server GraphQL implementations.

    \item \textit{Component definition}: In this step, we define the responsibilities of each component and detail their relationships. Three of the authors worked together in this step. Disagreements arising during this step were resolved by looking into various relevant sources such as documentation, source code, and usage. Fig~\ref{fig:graphql_ref_arch} shows our proposed architecture. Section~\ref{sec:results}  details all architecture layers and components of the current GraphQL ecosystem.
\end{enumerate}

\subsection{Topic Modeling}
Following previous work~\cite{blockchain, iot, security, mobile, bigdata, concurrency}, we perform topic modeling, specifically LDA, to study the GraphQL dataset we obtained from SO and identify the various GraphQL discussion topics. The following steps detail how we perform LDA.

\subsubsection{Importing StackOverflow Data Dump} \label{subsec:datadump}

% 1. why we chose stackoverflow (SO)
SO is a knowledge-based online website serving as a platform for software professionals and enthusiasts to discuss programming topics through questions and answers (Q\&As). Users of SO can ask questions, search for discussions about a certain topic, view discussions, search for specific tags, provide answers, accept answers, and vote (up or down) for answers. As of June 2024, it contains 24 million questions, 36 million answers, and 25 million users ~\cite{stackexchangesites}. Therefore, SO represents a great source of real-world software developer discussions to the extent that analyzing such a knowledge-based website could provide key insights to the relevant stakeholders of the respective domain communities, especially software practitioners.

To obtain the SO dataset, we download the December 2022 SO data dump and import the XML files into a relational MySQL database to facilitate query and analysis. The datasets for all the StackExchange sites are publicly available in the \textit{Internet Archive} under the \textit{Creative Common License}~\cite{internetarchive}. Each dataset contains seven XML documents (i.e.,  i) {\tt badges.xml}, ii) {\tt comments.xml}, iii) {\tt posts.xml}, iv) {\tt post\_history.xml}, v) {\tt post\_links.xml}, vi) {\tt tags.xml}, vii) {\tt users.xml}, and viii) {\tt votes.xml}). The {\tt posts.xml} file is what we are mainly interested in analyzing since that is where the developers' questions and answers data resides. Each post contains the following key data. {\tt Id} represents a unique identifier of a given post. {\tt PostTypeId} is a classifier to help indicate whether a given post is either a question or an answer. {\tt Title} represents the subject of the post, whereas {\tt Body} contains the developer's description of their post. {\tt AnswerCount} keeps track of how many answers the developer community has provided for a particular question. At the same time, {\tt ViewCount} represents the number of times a given post has been viewed. {\tt AcceptedAnswerId} is an identifier for the accepted answer for a given question and is set to {\tt null} for answers.

\subsubsection{Identifying GraphQL tags} 
\label{subsec:tags}
Once we have obtained and formatted the dataset, we filter it based on a tagset derived from a similar approach used in prior studies~\cite{concurrency,bigdata,iot}. This approach starts with an initial set of four tags ($T_{initial}$). Then, we identify tags that co-occur ($T_{candidate}$) with these initial tags on SO. Finally, we conducted an empirical study using two metrics defined in prior studies~\cite{iot,bigdata,concurrency,mobile} to identify an additional 56 relevant tags from the  $T_{candidate}$ set. Our final set includes 60 tags ($T_{final}$). We detail our approach to identifying GraphQL-related tags in the following.

\begin{enumerate}
    \item \textit{Initial tags:}  We started with \code{graphql} as the seed tag. We manually inspected the related tags listed on SO's tag description page ( \url{https://stackoverflow.com/questions/tagged/graphql}) and identified three additional tags for our initial set, which are \code{apollo}, \code{apollo-client}, and  \code{apollo-server}. Hence, $T_{initial}$ includes four tags.

     \item \textit{Identify candidate tags:} We search the SO data dump ($\mathcal{D}$) to identify a total of 21,137 questions ($\mathcal{P}$) that include at least one of the four initial tags. Next, we construct a set of candidate tags ($T_{candidate}$) by extracting tags of questions in $\mathcal{P}$. This step found a total of 2,534 candidate tags.
     
     \item \textit{Prepare final set of tags:} We use the two metrics, significance ($\mu$) and relevance ( $\nu$) used in prior studies~\cite{iot,bigdata,concurrency,mobile} to identify highly relevant tags. For a tag (t), these two metrics are measured as follows.
     
\vspace{2pt}
     Significance, $\mu$ (t) = $\dfrac{\# ~of~Questions~with~tag~t ~in~ \mathcal{P}}{\#~of~Questions~with~tag~t~in~\mathcal{D}}$

\vspace{2pt}

     Relevance, $\nu$ (t) = $\dfrac{\# ~of ~Questions ~with ~tag ~t ~in ~\mathcal{P}}{\# ~of ~Questions ~in~ \mathcal{P}}$
\vspace{2pt}

 We computed $\mu$ and $\nu$ for all the 2,534 candidate tags. A tag $t$ is significantly relevant to the \code{graphql} if $\mu$ (t) and $\nu$(t) are higher than some empirically determined thresholds. To identify an optimal threshold pair, we conduct an empirical study.  This step identifies a tag subset based on a value pair. Then, we manually inspect each tag to determine whether it is relevant to \code{graphql} by examining its description and recent questions listed on SO's tag description page. For a pair ($\alpha$, $\beta$), we also measure precision, which is defined as follows:

 Precision ($\alpha$, $\beta$) = $\dfrac{\# ~number ~of ~relevent ~tags}{\# ~number ~of ~tags ~selected~ based ~on ~(~\mu = ~\alpha ~\& ~\nu =~\beta) }$
\vspace{4pt}

 While setting higher threshold values yields higher precisions, we may miss some relevant tags. On the other hand, lowering thresholds decreases precision. Therefore, similar to prior SO studies, we experimented with a total of 49 combinations of ($\mu$, $\nu$)  from $\{0.01, 0.05, 0.1, 0.15, 0.2, 0.25, 0.3 \}$ $\times$ $\{ 0.001, 0.005, 0.01, 0.015, 0.02, 0.025, 0.03 \}$. 
 We found (0.2, 0.001) providing the set of relevant tags with a precision score of 96.7\%. These threshold values are similar to the ones used in prior studies~\cite{iot,bigdata,mobile,concurrency}. We consider this threshold pair optimal since changing either of the values significantly lowers precisions or missing relevant tags. %For example, with (0.15, 0.001), we obtain five additional tags, where four tags are irrelevant. On the other hand, with (0.2, 0.005), 34 tags from T$_{final}$ are not selected.
\end{enumerate}

At the end of these steps, we identified an additional 56 graphql-related tags. Table \ref{table:tags} shows 60 tags belonging to $T_{final}$. We also manually grouped the tags into five categories.

\begin{table}[ht]
\caption{A categorization of all 60 tags we identified to filter the SO data dump to only include GraphQL posts.}
\centering 
\resizebox{\linewidth}{!}{
\begin{tabular} {| p{0.26\linewidth}| p{0.72\linewidth}|}
\hline
\textbf{GraphQL Tag Category} & \textbf{GraphQL Tag Set} \\
\hline
 
GraphQL Specification (4) & graphql,  graphql-mutation, graphql-schema, graphql-subscriptions \\
\hline

GraphQL reference implementation (13) & apollo, apollostack, apollo-client, apollo-android,  apollo-angular, apollo-boost, apollo-cache-inmemory, apollo-federation,  apollo-ios,  apollo-server,  vue-apollo, react-apollo,     react-apollo-hooks \\ 
\hline

\hline
GraphQL programming-language implementations (8) & graphql-js, graphql-java, graphql-ruby, graphql-php, graphql-dotnet, graphql-go, graphene-django, graphene-python
  \\
\hline

GraphQL Tools \& Frameworks (34) & absinthe, aws-appsync,  ariadne-graphql, dgraph, gatsby, gatsby-image, gqlgen, grandstack, graphcool, express-graphql, prisma-graphql,   typegraphql,   graphql-tools, graphiql, graphql-codegen, graphql-spqr, hasura, hotchocolate, laravel-lighthouse,  netflix-dgs, nexus-prisma, postgraphile, prisma, prismic.io,  react-relay, relay, relayjs, relaymodern,  sangria, 
 wp-graphql, graphql-playground, neo4j-graphql-js,  graphql-tag, urql \\
\hline

GraphQL APIs (1) & github-graphql \\
\hline
\end{tabular}
}
\label{table:tags}
\end{table}

\subsubsection{Identifying GraphQL related posts} 
We searched our SO dump for all the posts with at least one of the 60 tags from $T_{final}$ and obtained 32,095 questions. Our search of the SO dump also found 31,195 answers to those questions. However, only 38.3\% of those answers are marked as `Accepted.' Following the recommendations of prior works~\cite{iot,mobile,bigdata,concurrency},  we only consider questions and accepted answers since unaccepted answers may include noises. Therefore, our topic modeling dataset includes 44,060 posts (32,095 questions and 11,965 accepted answers).

\subsubsection{Text preprocessing}
Before applying LDA to identify topics, we follow a nine-step preprocessing approach on posts to clean data. The following list describes our preprocessing steps.
 
\begin{enumerate}
    \item \textit{Conversion to lowercase}: We convert all the words to lowercase to ensure that the LDA model does not consider the capitalization of a word as a separate word from its lower-case counterpart.
    
    \item \textit{HTML removal}: We remove all the HTML tags since those are not relevant to discussion topics.
    
    \item \textit{URL removal}: Both questions and answers may include URL references (e.g., API reference) supporting the post. Since URLs do not communicate discussion topics, we remove all URLs from each text. 
    
     \item \textit{Stopword removal}: Since stopwords (usually non-semantic words such as articles, prepositions, conjunctions, and pronouns) are irrelevant to identifying topics, we remove all stopwords from each text.
     
     \item \textit{Domain-specific term removal}: We exclude GraphQL domain-specific terms (e.g., "graphqL", "apollo", "client", "server"), which are frequent among various discussion topics. Additionally, we exclude exemplary schematic terms that our experience shows do not add significant value to the discovered topics. This set includes variable names such as "x," "foo," "bar," and "baz." Collectively, these exclusions help reduce noise, improve stability, and increase topic coherence.
    
    \item \textit{Number/punctuation removal}: since numbers and punctuation symbols are not tied to a particular topic, we remove those.
    
    \item \textit{Stemming}: this process reduces words to their origins by removing suffixes. For example, `mining', `mine'    and `mined' would all become `mine.' We use the Snowball Stemmer~\cite{snowballstemmer} to convert derived words to their base form. 
    
    \item \textit{Bigram detection:} a bigram is a pair of consecutive words frequent in a corpus. We identify the bigrams that appear in at least 20 posts and merge those into a single token using an underscore (`\_'). For example, `public key', becomes 'public\_key'.
    
    \item \textit{Exclusion of very high or low-frequency words}: we exclude high frequency (i.e., words appearing in more than 20\% documents, since those may not be tied to any particular topic. It also excludes low-frequency words (i.e., words appearing in less than ten documents) since those are not significant terms for topic analysis but can increase the model training time.
\end{enumerate}

\subsubsection{Determining Number of Topics} 
\label{sec:topic-count-study}
\begin{figure}[ht]
  \centering
  \includegraphics[width=0.75\textwidth]{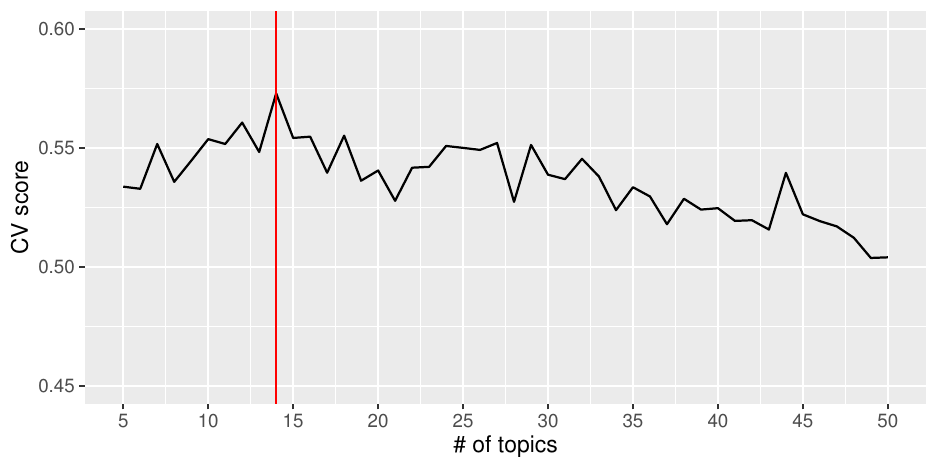}
  \caption{ Best $c\_v$ score Vs.  \# number of topics (K). We repeated LDA model training ten times for each value K and took the best CV score.  }
  \label{fig:optimum-topic}  
\end{figure}

\begin{figure}[ht]
  \centering
  \includegraphics[width=0.75\textwidth]{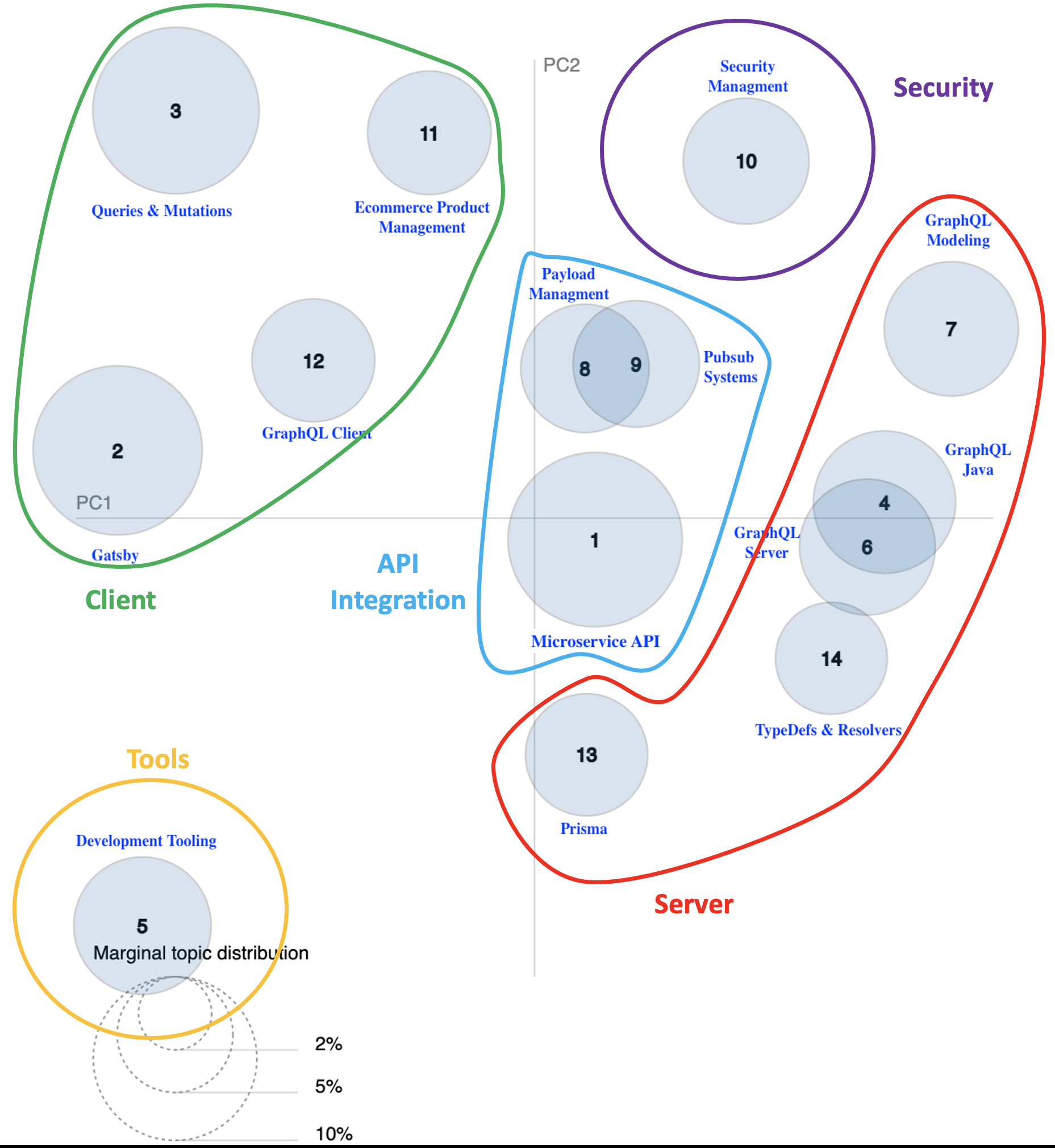}
  \caption{Intertopic distance map showing topic distribution, sizes, and overlap at the number of topics K = 14 (less overlap is better)}
  \label{fig:inter_topic_distance_map}  
\end{figure}

The optimal number of topics ($K$) is essential to avoid undesirable outcomes. For example, Mehdi \textit{et al.}~\cite{bigdata} documented that choosing a small K value may result in multiple, possibly unrelated, topics combined into one. On the contrary, a large value of K may ultimately split what would otherwise be a single coherent topic into multiple subtopics. Therefore, we run 45 LDA experiments varying the range of K between 5 and 50 with a step increment of 1. For our LDA implementation, we leverage the Gensim LDA Python library~\cite{gensim}. Our configuration brute forces the number of iterations for all experiments at 500 and 1000 iterations following previous work~\cite{bigdata, iot}.
Additionally, Bigger \textit{et al.}~\cite{bigger} showed that hyperparameters have little impact on the accuracy of the produced topics; therefore, we set them at their default settings. To determine the best number of topics, we employ three metrics: topic coherence, topic stability, and inter-topic distance map. We measure topic coherence via the $c\_v$ score~\cite{roder2015exploring}, extracting the top 15 words from each topic. For topic stability, we utilize the Jaccard similarity score by repeating every experiment 10 times and comparing the top 10 words of the produced topics.  Lastly, we use the interactive topic model visualization tool (LDAvis) proposed by Sievert \textit{et al.}~\cite{ldavis} to visually inspect the size, distance, and overlap of the discovered GraphQL topics.  We take the best $c\_v$ score produced over ten runs for each value of K and also store a classification of all documents using the best model. Figure~\ref{fig:optimum-topic} shows variations of CV scores with the number of topics.
Our experiments show that (Figure~\ref{fig:optimum-topic}), at 500 iterations, K = 14 scores the highest $c\_v$ measure (0.57) with a stability score of (0.92). 
The inter-topic distance map shown in Fig~\ref{fig:inter_topic_distance_map} shows little visual overlap across the generated topics. Therefore, we conclude that K = 14 provides an optimal model for this study's goal.

\subsection{Topic evolution and trends}
During our experiment to determine the optimum number of topics for our corpus (Section~\ref{sec:topic-count-study}), we identified a set of K topics $(z_1; . . . ; z_K)$ and a set of topic probability vectors for each post ($p_i$) in our corpus. Following the recommendations of  prior studies~\cite{blockchain,han2020programmers,barua2014developers}, we compute the probability of a post ($p_i$) belonging to the topic ($z_k$) as $ \theta(p_i, z_k)$, where $\sum_1^{K} \theta(p_i, z_k) =1$.
With the topic probability vector for a post ($p_i$), we computed its dominant topic as 

\begin{equation}
    dominant (p_i) = z_k :  \theta(p_i, z_k) = max \{ \theta(p_i, z_j)\}, 1 \le j \le K
\end{equation}

Similar to prior studies~\cite{blockchain,iot}, we measure the popularity of a topic ($z_i$) in our dataset using the ratio of posts with $z_i$ as the dominant topic, as follows.

\begin{equation}
    popularity (z_k, c_j) =  \frac{|\{ d_i\}|} {|c_j|} : dominant (d_i) = z_k, 1 \le i \le c_j
\end{equation}

We use the following two metrics used in prior studies~\cite{blockchain,iot,han2020programmers} to measure topic trends over time. The first metric, called \emph{Topic absolute impact}, measures the absolute proportion of posts related to a particular topic $z_k$ in the month $M_i$. It is defined as: 
\begin{equation}
\label{eq:absolute_impact}
    impact_{absolute} (z_k, M_i) = \sum_{p_i \in  D(M_i)}   \theta(p_i, z_k)
\end{equation}
where $D (M_i)$ is the set of posts in month $M_i$. 
On the other hand, the \emph{Topic relative impact} metrics normalize the absolute impact by computing the ratio of posts in a month tied to a particular topic. It is computed as :
\begin{equation}
\label{eq:relative_impact}
    impact_{relative} (z_k, M_i) = \frac{1}{|D(c, M_i)|} \sum_{p_i \in  D(M_i)}  \theta(p_i, z_k)
\end{equation}
where $|D(c, M_i)|$ is the set of all posts in the month $M_i$.

\subsection{Topic hierarchy} 
 We assign each post to a topic cluster based on its dominant topic ($z_k$).  
   We ran a second-level LDA experiment with each topic cluster's posts to discover fine-grained subtopics for each topic separately. Our second-level topic modeling follows a strategy similar to our top-level LDA experiments. However, as the number of documents is smaller, we varied $K_2$ from 2 to 5. This range choice ($K_2$) was motivated based on prior studies, where the number of identified subtopics varied between 2 to 5~\cite{bigdata,iot}.   We empirically identified the  $K_2$ value producing each cluster's highest $c\_v$ score. At the end of this second-level experiment, we identified two to five subtopics for each first-level topic.

 \subsection{Topic-Naming Strategy}
\label{topic-naming}
To name our discovered GraphQL topics, we adopt the following approach. Initially, we analyze the discovered top terms of each topic utilizing the interactive topic visualization tool LDAvis~\cite{ldavis}. 
We utilize the relevance metric (i.e., \textit{Lambda}) to identify the top-30 terms. 
At \textit{Lambda=0}, a set of top-30 terms contains the most relevant words but not necessarily the most frequent, whereas \textit{Lambda=1} favors the most frequent words over their corresponding relevance. We choose \textit{Lambda=0.6} because it is the recommended value according to the authors of LDAvis~\cite{ldavis} for visualization and selection of relevant terms.  

Three authors independently reviewed each topic's terms and subtopic terms produced through second-level LDA to assign it a name and suggest their architectural mapping, which we will elaborate on in Section~\ref{topic_architecure_mapping_methodology}. During this process, they also reviewed sample questions where the topic under review was dominant. After this independent name assignment process, they met to compare and contrast their assigned names. While the names assigned to topics differed due to differences in terminologies, in most cases, assigned names were congruent. They discussed these name assignments, reviewed more questions as necessary, and determined an agreed-upon name for each topic. Finally, they randomly sampled 20 SO posts from each topic to validate/refine their naming.  This process required multiple discussion sessions due to the sizeable number and amount of topics, their associated subtopics, and architectural mapping.

\subsection{Topic to architectural layer mapping} 
\label{topic_architecure_mapping_methodology}

In this step, we map each of the identified topics to the architecture layers of the GraphQL ecosystem developed by us. 
As a starting point, we manually analyze the discovered GraphQL topics to determine their respective mapping to the architecture layers. For example, a topic dominated by discussions about the Prisma~\cite{prisma} toolkit likely maps to the server side of the architecture rather than the client side because Prisma is a tool that aims to migrate existing databases to support a GraphQL API instantly. We found mapping ten of the 14 topics straightforward since they cohesively contain posts constituting well-defined ecosystem concepts. 
On the other hand, the remaining four topics, which include \textit{Pubsub Systems}, \textit{GraphQL Java}, 
 \textit{Payload management} and  \textit{Microservice Architecture} contain posts that span a broad spectrum of cross-cutting concepts.  Therefore, it is not easy to map such topics to one layer.

To simplify our analysis, we consider a topic mapped to a particular layer if the posts related to a particular layer form most of the posts assigned to a topic by our first-level LDA model. We randomly selected 30 posts for manual analysis to identify the dominant layer for each of these four topics. Our analysis found three of the four topics being mapped to the `API Integration' layer. This mapping also justifies the cross-cutting nature of questions belonging to these three topics. These three topics are also positioned in the center of our intertropical distance map shown in Figure~\ref{fig:inter_topic_distance_map}, indicating their cross-cutting characteristics. Section~\ref{sec:results} elaborates on the concepts in all topics and details their respective architecture layers.

\subsection{Analysis of popularity and difficulty}

Prior studies have used three metrics to compute the popularity of topics~\cite{iot,bigdata,al2024far}, which are i) the average number of views, including both registered users and visitors of SO, ii) the average score, a metric to indicate the number of times a post has been up or down voted, and iii) the average number of questions of a topic that are marked by users as a favorite question.

We can not use the average favorite count since this metric is no longer present in the recent SO data dump. Moreover, we use medians instead of averages for the remaining two measures since the results of our Anderson–Darling tests~\cite{anderson1952asymptotic} indicate that those significantly differ from a normal distribution. 
We measure the difficulty of a topic using two metrics used by prior studies: i) the percentage of questions of a topic that have no accepted answers and the median time needed for questions of a topic to receive accepted answers~\cite{iot,bigdata,al2024far}. Intuitively, a topic with fewer accepted answers received or requiring a longer time is more difficult. %Similar to prior studies~\cite{al2024far,bigdata}, 

\section{Results} \label{sec:results}

The following subsections present the results of our five research questions, respectively.

\subsection{RQ1: How can we define a referenced architecture of the current GraphQL ecosystem?} 
\label{results_arch_layer}
\begin{figure}[ht]
  \centering
  \includegraphics[width=15cm]{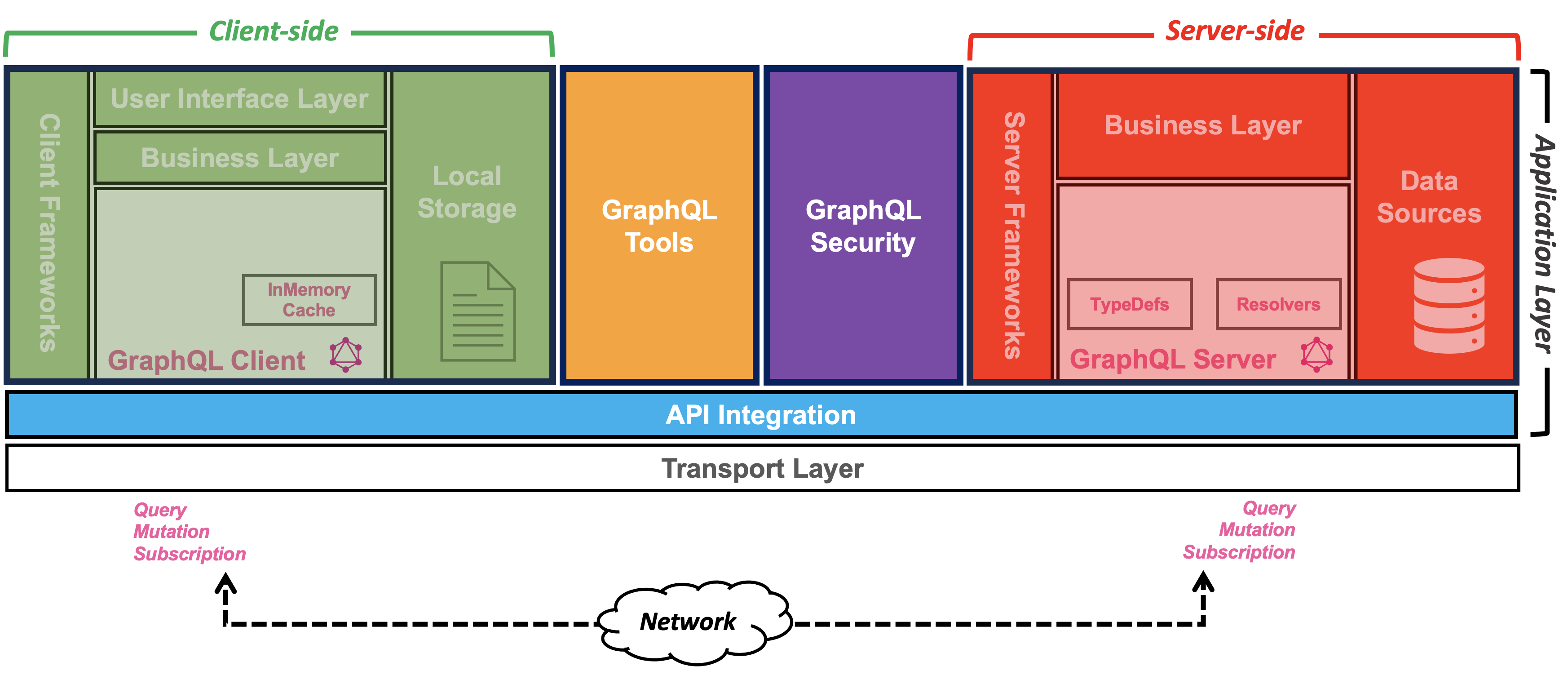}
  \caption{Our proposed reference architecture of the GraphQL ecosystem.}
  \label{fig:graphql_ref_arch}  
\end{figure}

Figure~\ref{fig:graphql_ref_arch} represents our reference architecture of the GraphQL ecosystem resulting from the reference architecture derivation process described in ~\ref{methodology:derive_ref_architecture}. The diagram depicts a client-server architecture where the GraphQL client and server communicate over a network via an API integration layer. In a client-server architecture, a GraphQL client requests resources from a GraphQL server using queries, mutations, or subscriptions. The server, in turn, resolves requests, performs security checks and business validations, fetches resources from data sources, and responds accordingly. The client receives the GraphQL response, stores it in local storage or cache if necessary, performs respective business functionality, and renders the data on the user interface.

As shown, both GraphQL client and server operate at the application layer of the Open Systems Interconnection (OSI) model. The GraphQL server has four components: business, data sources, GraphQL server, and framework. The GraphQL client has five components: user interface, business, GraphQL client, framework, and local storage. The diagram also shows that the tools and security components are cross-cutting because they apply to multiple GraphQL client or server layers. The following subsections detail all layers and components of our reference architecture for the GraphQL ecosystem.

\subsubsection{Server side}
According to our reference architecture, the four components of the server side of GraphQL ecosystems are:

\begin{enumerate}
    
    \item \textbf{Business Layer} sits atop the layers of the GraphQL server. It is responsible for defining domain-specific business rules and executing and validating business logic according to such rules.
    
    \item \textbf{GraphQL Layer} parses, validates, and resolves client GraphQL requests (i.e., queries, mutations, and subscriptions). It contains schema-type definitions (i.e., typedefs) and resolver functions. The former defines object types that correspond to the GraphQL schema. It allows the server to validate incoming GraphQL requests. In contrast, the resolvers are functions that allow the server to handle the GraphQL operations and fetch the resources from various data sources before responding to the client.

    \item \textbf{Data Sources} represents sources that GraphQL resolver functions connect to fetch, populate, and serve data to the requesting clients. Data sources may be external, such as database systems and RESTful APIs, or internal, such as in-memory data. Moreover, data sources are not only limited to the GraphQL layer since the business or other layers may connect to them directly.

    \item \textbf{Framework} represents libraries and frameworks developers utilize to support GraphQL in their server-side applications across various programming languages. Popular frameworks include Graphene-Django for Python, GraphQL-Java for Java, and Express for Apollo-server. The full list of frameworks is available on the GraphQL website~\cite{graphqltools}.
\end{enumerate}

\subsubsection{Client side}
The five components of the client side of GraphQL ecosystems are as follows.
\begin{enumerate}

    \item \textbf{User Interface Layer} sits at the top of the client-side layers. It applies to web-based front-end GraphQL clients such as web and mobile applications. It incorporates graphical user interfaces (GUIs) and technical web view components to visually present data to end-users and allow user interaction.  
    
    \item \textbf{Business layer} has a responsibility similar to that of the server side, where clients perform various business functionality for their respective front-end applications. 
    
    \item \textbf{GraphQL layer} allows a GraphQL client to create various GraphQL operations to interact with the GraphQL server. A client creates, validates, and sends requests to the GraphQL server. Then, it receives responses, caches results, and maintains the local state. GraphQL clients also utilize middleware to translate requests and responses to the proper application protocol and manage authentication. GraphQL clients vary in terms of features they provide, ranging from basic features to fully provisioned, such as Apollo, which provides pagination and normalized caching.
    
    \item \textbf{Local Storage} refers to methods available in a web browser or mobile application that allow them to store data locally on the user's device. This data is persisted across browser sessions and is accessible after a browser or mobile application is closed and reopened. Local storage is primarily used for saving preferences, caching data to improve the performance of a GraphQL client, or allowing for offline use to reduce constant dependence on the GraphQL server.
    
    \item \textbf{Framework} is similar to the framework component on the server side where software developers utilize libraries and frameworks to support their front-end GraphQL applications. React is one of the most popular frameworks used, and the full list is available on the GraphQL website~\cite{graphqltools}.
\end{enumerate}

\subsubsection{API Integration}
This layer has a few key responsibilities, as follows. Since the GraphQL specification is by design transport-layer-agnostic~\cite{graphql_spec}, GraphQL client/server implementations utilize an API integration layer responsible for managing lower-level requests/responses from underlying application programming interfaces (APIs) and protocols. HTTP is the most widely used protocol for request-response models, whereas websockets~\cite{websockets} is the most prevalent for publish-subscribe models. Additionally, this layer decouples the GraphQL layer from the formats of integrated frameworks using middleware or integrations. The Apollo reference implementation lists community-maintained integrations such as AWS Lambda, Fastify, Koa, and Hapi. The full list is available in the Apollo integration documentation~\cite{apollo_server_integrations}.
Moreover, one premise in GraphQL is the usage of a single GraphQL endpoint (typically \lstinline{/graphql}) to serve all requests. Mandating all services be exposed via a single API endpoint for large-scale organizations is neither scalable nor efficient~\cite{graphqlfederation}. Organizations employ GraphQL API gateway or federation to overcome such challenges, which fundamentally allows splitting the GraphQL schema into smaller independent subgraphs that may be developed separately. The most popular GraphQL federation implementation is Apollo Federation~\cite{apollofederation, graphqlfederation}. In GraphQL federation, an API Gateway has a crucial role of receiving client requests via a single GraphQL endpoint and routing them to the appropriate GraphQL subgraph, thereby stitching all subgraphs and creating a supergraph of an organization's overall GraphQL schema. Lastly, it provides a single enforcement point to centralize specific security concerns such as authentication, API key management, and rate limiting.

\subsubsection{GraphQL Security}
GraphQL APIs present unique security challenges because they are flexible compared to REST APIs. While one strength is that clients can form complex queries to fetch exactly the data they require, this also introduces specific security considerations to prevent malicious queries, protect sensitive data, and ensure the data integrity and availability of the server. Therefore, the security component in the GraphQL ecosystem is responsible for defining strategies, including input validation and sanitization, analyzing query complexity, limiting query depth, rate limiting, authentication, and authorization.

\subsubsection{GraphQL Tools}
This component primarily encompasses the various GraphQL development tools, plugins, and IDEs developers employ to manage their GraphQL applications throughout the software development lifecycle. This includes tools to generate code, build, package, deploy, install, test, debug, and introspect GraphQL APIs. For example, developers may use code generation tools to autogenerate GraphQL schema, code, or documentation. Code generation tools that adopt schema-first approaches allow developers to design GraphQL schema, and the tools generate the programming-language-specific code corresponding to their GraphQL type definitions. On the other hand, code-first tools imply that developers write code before tools auto-generate the respective GraphQL schema. Examples of artifact life-cycle management tools include NPM, Apache Maven~\cite{maven}, and Gradle~\cite{gradle}. To introspect, test, and debug GraphQL APIs, software developers may use various tools such as GraphiQL, Apollo Studio, Insomnia, NPM, postman, curl, and GraphQL Playground. An elaborate list of tools can be found on the GraphQL website~\cite{graphqltools}.

\begin{boxedtext}
      \textbf{Takeaway 1.} GraphQL ecosystem has five primary architecture layers and is complex with various interconnected tools, technologies, and platforms from those categories. 
  \end{boxedtext}

\subsection{RQ2: What do GraphQL developers discuss on StackOverflow?}
\label{sec:generate_topics}
\begin{table}[ht]
\centering
\caption{A list of our 14 discovered GraphQL topics, including our proposed topic names and their respective top-10 stemmed words.}
\label{table:topics}
 \begin{tabular}
 {| p{0.02\linewidth} | p{0.25\linewidth}| p{0.65\linewidth}|}
\hline
\# & \textbf{Topic Name} & \textbf{Top-10 Terms} \\
\hline
1 & \textit{Microservice API} & servic comment rest endpoint databas singl librari multipl backend support
\\
\hline
2 & \textit{Gatsby} & imag gatsbi content site plugin layout blog img tag style
\\
\hline
3 & \textit{Queries \& Mutations} & compon load render react prop usequeri usest button state useeffect
\\
\hline
4 & \textit{Java Implementation} & public  class date entiti private defin column scalar filter schema
\\
\hline
5 & \textit{Development Tooling} & npm build version gatsbi instal depend command modul npm\_warn fail
\\
\hline
6 & \textit{GraphQL Server} & resolv schema await role apolloserv access ctx promis permiss datasourc
\\
\hline
7 & \textit{GraphQL Modeling} & model relat class def map refer float self graphen line
\\
\hline
8 & \textit{Payload Mgmt} & fragment upload relay cursor enum edg pagin kind viewer direc
\\
\hline
9 & \textit{Pubsub Systems} & subscript updat subscrib messag event appsync websocket connect lambda pubsub
\\
\hline
10 & \textit{Security Mgmt} & email usernam token login password await userid  profil regist account
\\
\hline
11 & \textit{Ecommerce Product Mgmt} & product categori slug order index filter test array mock key
\\
\hline
12 & \textit{GraphQL Client} & cach typenam header inmemorycach httplink apollocli cooki link token appolloprovid
\\
\hline
13 & \textit{Prisma} & prisma databas connect migrat db postgr prismacli port docker environ
\\
\hline
14 & \textit{GraphQL Resolvers} & graphqlstr id graphqlobjecttyp resolv arg graphqllist graphqlnonnul mongoos schema express
\\
\hline
 \end{tabular}
\end{table}
\begin{table}[ht]
\centering
\caption{A list of the names of our 14 first-level topics, the number of their associated second-level subtopics, and the subtopic names}
\label{table:subtopics}
 \begin{tabular}
 {| p{0.2\linewidth}| p{0.15\linewidth} | p{0.55\linewidth}|}
\hline
\textbf{Topic Name} & \textbf{\# of Subtopics} & \textbf{Subtopic Names} \\
\hline
\textit{Microservice API} & 4 & Service API, Client Libraries, Database, Schema \& Resolvers
\\
\hline
\textit{Gatsby} & 4 &  Image, Plugins, Script, Slug
\\
\hline
\textit{Queries \& Mutations} & 3 &  Data Fetching, Form Submission, Gatsby Prefetching
\\
\hline
\textit{Java Implementation} & 4 &  Scalar Types, Dependency Mgmt, Java Constructs, ORM
\\
\hline
\textit{Development Tooling} & 2 &  Build Tooling, Deploy Tooling
\\
\hline
\textit{GraphQL Server} & 5 & 
Server Management, Security, Modeling,
Database, Resolvers
\\
\hline
\textit{GraphQL Modeling} & 2 &  Graphene-Django, Prisma
\\
\hline
\textit{Payload Mgmt} & 5 &  Pagination \& Filtering, Mime-type Mgmt, JSON Content, GraphQL Fragments, Enum Types
\\
\hline
\textit{Pubsub Systems} & 4 &  Apollo Client, Websockets, Appsync, Event Modeling
\\
\hline
\textit{Security Mgmt} & 3 &  User Authentication, User Payload, User Model
\\
\hline
\textit{Ecomm Product Mgmt} & 2 &  Product Testing, Product Selection
\\
\hline
\textit{GraphQL Client} & 4 &  User Interface, Apollo Links, State Mgmt, Cache Mgmt
\\
\hline
\textit{Prisma} & 2 &  Schema Migration, Deployment (Docker)
\\
\hline
\textit{GraphQL Resolvers} & 3 &  TypeDefs, Mongoose, NodeJS
\\
\hline
 \end{tabular}
\end{table}
 After conducting our LDA experiments and following our topic naming and architecture mapping strategies, we summarize the results of our 14 discovered GraphQL topics in Tables~\ref{table:topics} and ~\ref{table:subtopics}. The former table lists each topic's index (I) mapping to its intertopic distance map in Fig~\ref{fig:inter_topic_distance_map}, our proposed topic name, and a list of its corresponding top-10 words. Table~\ref{table:subtopics} lists the names of all 14 topics as well as the names of their subtopics. Additionally, we formally define all the topics we have discovered and provide evidentiary sample posts from the SO website to support our naming.

\begin{enumerate}

    \item \textbf{Microservice API:}
    This topic contains developer discussions about the GraphQL microservice API. A microservice architecture breaks business functions into small, independent, scalable, and loosely coupled services with well-defined APIs instead of one monolithic application. In GraphQL, the microservice design entails that services expose GraphQL APIs (as opposed to legacy APIs, e.g., REST) for clients to consume. Our second-level LDA experiments suggest four subtopics as follows: \textit{API Consumption}, \textit{API Fullfillment}, \textit{API Service}, and \textit{API Autogeneration}. The \textit{API Service} subtopic is concerned with designing GraphQL services, exposing GraphQL API endpoints, and utilizing API gateways. The prominent service API in this topic is Amazon AppSync~\cite{appsync}, part of the Amazon Amplify platform. The \textit{API Consumption} subtopic involves libraries that GraphQL clients utilize to consume GraphQL APIs, such as "relay," "react," "redux," and "Gatsby."  The \textit{API Autogeneration} subtopic concerns specific backend libraries that auto-generate GraphQL schema from existing databases to instantly expose GraphQL APIs such as Prisma~\cite{prisma}, Hasura~\cite{hasura} and PostGraphile. Lastly, the \textit{API Fullfillment} discusses resolver to fulfill client requests.  Therefore, we have mapped this topic to the \textit{API Integration} layer of the GraphQL ecosystem.
     An example post from SO is the following: \post{"My company runs a microservice architecture that has 50+ services that are powered by a single GraphQL endpoint, which orchestrate the calls among our services, powering our Android \& iOS applications for our end-users. We're in process of creating a new product that's going to not be used by those end-users, but for companies that offer goods for our end-users through our apps. TL;DR: things like showing performance data regarding their sales through our platform... In case your company had this discussion before, what was the final decision made, what was taken into consideration?"} [\hyperlink{https://stackoverflow.com/questions/54676109/should-i-have-multiple-graphql-instances-or-just-a-single-one}{$Q_{54676109}$}].
    
    \item \textbf{Gatsby}~\cite{gatsby}:
    This topic represents developer discussions about Gatsby, an open-source frontend framework that is based on the React framework and primarily targets static website generation. Gatsby also supports other rendering options and is backed by a powerful GraphQL data layer that allows for dynamic data fetching from multiple sources such as file systems, databases, REST, and GraphQL APIs (Data source). Such data layer's ability is enabled by various Gatsby plugins and automatic GraphQL schema inference tools (Framework). This topic has four subtopics: \textit{HTML, Images, Slug, and React Integration}, and \textit{image}. The \textit{HTML} subtopic corresponds to the \lstinline{<script>} tag in HTML, which Gatsby utilizes to help load JavaScript scripts. The \textit{React Integration} subtopic refers to the React-based Gatsby plugins developers utilize to build, link, style, extend, and load their Gatsby projects. The \textit{Slugs} subtopic contains Gatsby posts associated with dynamic page creation. A slug uniquely identifies a resource on the web with a user-friendly URL. The \textit{Image} subtopic is concerned with automatically optimizing and displaying images for the web, such as compressing and resizing them to reduce file sizes without sacrificing quality.
    Since all subtopics relate to client-side integration, we map it to the \textit{Client layer}.
    For example, a question on this topic asks, \post{"I have a Gatsby site set up with Sanity CMS for blogging. Current I am unable to properly target individual blog post (clicking on "First Blog" link will render content from "Second Blog"). My guess would be this is due to how graphQL query is setup in template/blog.js? If so, what changes would need to be made to properly target individual blog post?"} [\hyperlink{https://stackoverflow.com/questions/72665818/how-to-query-sanity-blog-post-with-gatsby}{$Q_{72665818}$}]. 

    \item \textbf{Queries \& Mutations}: This topic contains developer discussions about triggering, executing, and rendering GraphQL queries and mutations on the front end, especially using Apollo Client~\cite{apolloclient}. The discussions involve managing user interface forms and triggering GraphQL queries and/or mutation eagerly or lazily via the various hooks (e.g., \lstinline{useQuery}, \lstinline{useMutation}, and \lstinline{useLazyQuery}), fetch and load data in the local state of the GraphQL client, and render responses on the user interface. The React framework is the most dominant, as the word "react" is present in 76\% of the discussions. Our second-level LDA experiments produce three subtopics. The first subtopic is \textit{Form submission}, which includes top terms such as "button", "form", "submit", "field" and "label". The second subtopic is \textit{Data fetching}, which discusses various hooks to manage GraphQL queries and/or mutations from the React framework. It contains top words such as "useEffect", "useQuery", "useMutation" and "useLazyQuery". The last subtopic is about prefetching linked page's resources in Gatsby~\cite{gatsby} to provide a faster user experience. Based on the discussion areas of the three subtopics, we map it to the \textit{Client layer}.  The following is an example of a question posted on SO: \post{"I was going through the documentation of Apollo React hooks. And saw there are two queries hooks to use for which is useQuery and useLazyQuery. I was reading this page. https://www.apollographql.com/docs/react/api/react/hooks/. Can someone explain me what is the difference between them and in which case it should be used."}.  [\hyperlink{https://stackoverflow.com/questions/63681650/what-is-the-difference-between-usequery-and-uselazyquery-in-apollo-graphql}{$Q_{63681650}$}]. 
    
    \item \textbf{Java Implementation}:
    Posts on this topic concern implementing a GraphQL server in Java. Discussions include developers requesting recommendations on the best GraphQL-Java implementation libraries, modeling GraphQL using Java domain models, and defining GraphQL Java types. Developers also discuss underlying service containers such as Spring~\cite{spring}. Our second-level LDA experiments produce four subtopics: \textit{Scalar types, Dependency management, Java constructs} and \textit{ORM mapping}.   The \textit{scalar types} subtopic is concerned with managing custom scalar types in Java, especially date/time types. This includes the scalar type creation, serialization, format, and JSON representation. There are two types of dependency management techniques present in \textit{dependency management}, namely: bean and artifact dependencies. The former discusses resolving the runtime injection of Java beans, especially in the Spring framework~\cite{spring}, which GraphQL Java also uses as an application service container. The term "spring" is present in 13\% of the discussions. The artifact dependency involves resolving build time library dependencies using build life-cycle management tools such as Maven~\cite{maven} and Gradle~\cite{gradle}.  The \textit{Java constructs} subtopic contains discussions around Java classes, interfaces, and functions utilized to resolver GraphQL queries. Lastly, the \textit{ORM mapping} subtopic discusses the object-relational mapping of Java entities to database tables.   Based on our analysis of the posts and subtopics, we map this topic to the \textit{Server layer} of the reference architecture. 
    One example post is the following: \post{"I need a graphQL client for java spring application to communicate with another microservice based on graphQL API. I know about Apollo Android but it seems to be not implemented with maven...Any ideas and suggestions?"} [\hyperlink{https://stackoverflow.com/questions/68019255/graphql-client-for-java-application}{$Q_{68019255}$}].

    \item \textbf{Development Tooling}:
    As the top 10 words of this topic suggest, it contains questions and answers concerning tools developers utilize through the development of GraphQL projects. These tools play a crucial role in the GraphQL ecosystem by facilitating the installation and management of libraries, tools, and dependencies necessary for both server-side and client-side GraphQL development. We map this topic to the \textit{ Development Tools} layer of our reference architecture. A few examples of tools present in developer discussions of this topic include Yarn, npm, and Webpack~. Node Package Manager (NPM) tops the top 10 terms of the parent topic and is present in 36\% of the discussions. Therefore, we consider it the most widely used artifact package manager in the context of GraphQL. Our second-level GraphQL experiments produce two subtopics, namely: \textit{Build tooling} and \textit{Deploy tooling}. As the names suggest, \textit{Build tooling} contains discussions that concern resolving dependencies or building GraphQL packages, whereas \textit{Deploy tooling} concerns deploying GraphQL artifacts to cloud platforms such as Netlify and Heroku.  The following example SO answer suggests a developer use a command-line tool that utilizes NPM to generate a GraphQL project with support for typescript: \post{"I am using a GraphQL CLI. You would install it like so \lstinline{npm install -g graphql-cli} then generate your GraphQL project with TypeScript support"} [\hyperlink{https://stackoverflow.com/questions/50905873/apollo-graphql-server-typescript}{$Q_{50905873}$}].

    \item \textbf{GraphQL Server}:
    In this topic, software developers have various discussions about managing GraphQL servers and naturally map to the \textit{Server layer} of our reference architecture. The spectrum of discussion points includes creating, configuring, and securing the server in addition to managing server middleware, business modeling, and data access. The most prominent server implementation developers discuss is Apollo Server which occurs in 34\% of the discussions. Our second-level LDA experiments produce five subtopics as follows: \textit{Server management, Security, Modeling, Database} and \textit{Resolvers}. The first subtopic concerns managing the GraphQL server (e.g., Apollo Server), HTTP middleware such as the \textit{Express} server, API endpoints, and the federation gateway. The \textit{Security} subtopic contains discussions about \textit{Auth Guards}, which is a mechanism to protect certain parts of the GraphQL API by ensuring that the user making a request is authenticated and authorized to perform the requested operation. It also involves the management of users, security groups, and permissions. In the \textit{Modeling} subtopic, discussions are regarding managing business data models for GraphQL resources. Posts in the \textit{Database} subtopic involve managing database tables and various selection, filtering, and join queries to resolve GraphQL requests. Lastly, the \textit{Resolvers} subtopic concerns how requests are resolved and how data is retrieved, processed, and returned in a GraphQL API. In this subtopic, there is a special emphasis on passing input arguments and resolving nested queries. An example post from SO is: \post{"I want to implement cursor based pagination in Apollo graphql server. I have prepared schema with pagination requirement. But i am stuck at resolver side. Here is my schema... Is it possible to resolve in resolvers? If yes, can any one please tell me how to implement it"} [\hyperlink{https://stackoverflow.com/questions/44383522/pagination-in-apollo-graphql-server}{$Q_{44383522}$}].  
    
    \item \textbf{GraphQL Modeling}:
    This topic contains questions and answers about GraphQL schema definition and its respective object relationship modeling. It aims to generate GraphQL schema definitions and realize their equivalent object model or entity model in a programming language or database schema, respectively. Since data models are implemented on the server side, we map this topic to the \textit{Server layer} of our reference architecture.  Our second-level LDA experiments suggest two subtopics that correspond to two backend framework models. The first framework is Prisma, which we detail separately in the Prisma topic description. The second is Graphene-Django~\cite{graphenedjango}, a Python framework to enable the development of GraphQL backend APIs. As the name suggests, Graphene-Django represents an integration of two Python frameworks, namely Graphene and Django. The former is an open-source Python library that allows for building GraphQL APIs using a code-first approach, contrasting with schema-first approaches. For example, in the Ariadne~\cite{ariadne} python library, developers start with writing the GraphQL schema before the respective code is generated thereafter. On the other hand, Django~\cite{django} is an open-source Python web framework for building server-side projects and web pages. Graphene-Django supports query validation, subscriptions, middleware, and data modeling backed by data loaders to facilitate integration with various data sources. The following is an example question from SO: \post{"I am using Django 4 with graphene-Django 3 and I need to structure a response that must be 4 levels of lists and the bottom level a dictionary...GraphGL takes more than 30s to parse the given result...I am not an expert in GraphGL...how could I speed up graphene-Django to parse this response faster..."} [\hyperlink{https://stackoverflow.com/questions/74154000/improve-performance-of-graphql-graphene-django-for-4-levels-of-nested-lists}{$Q_{74154000}$}]. 

    \item \textbf{Payload Management}: Questions and answers in this topic are regarding either constructing valid GraphQL requests (e.g., fragments) or consuming response messages appropriately. Our second-level LDA experiments suggest five subtopics, namely: \textit{Pagination and Filtering, Mime-type management, JSON, Fragments}, and \textit{Enum types}. The first subtopic involves managing search operations to perform pagination, cursor navigation, selection, filtering, sorting, and ordering using the \textit{Relay} framework. This framework provides complete specifications for GraphQL cursor connections and pagination. The \textit{Mime-type management} subtopic contains posts concerning the management of multipurpose internet mail extensions, such as uploading files and images. The term "file" appears in 22\% of the parent topic's posts. In the \textit{JSON} subtopic, developers inquire about clients receiving, parsing, validating, converting, and formatting JSON payloads. We note that the term "json" is present in 19\% of the posts. We also note that developers ask questions regarding handling erroneous messages or correcting ill-formed GraphQL requests and validating them against the GraphQL schema. In the last two subtopics, developer discussions are about managing client-side \textit{Fragments} and passing \textit{Enum types} as part of the GraphQL requests. For instance, the term "error" appears in 32\% of the posts. Based on our analysis of the subtopics, we map this topic to the \textit{API Integration} layer. 
    The following question is an example: \post{"Our app uses Algolia for searching which allows for two typical forms of pagination: by page page=3\&hitsPerPage=5 or by offset offset=10\&length=5 GraphQL preferences cursor-based pagination where a recordID and length are provided friendsConnection(first:5 after:"Y3Vyc29yMQ==") Is there anyway to translate info from a cursor-based pagination request into a page or offset pagination request?"} [\hyperlink{https://stackoverflow.com/questions/50033816/graphql-cursor-based-pagination-converted-into-page-or-offset-pagination}{$Q_{50033816}$}].
    
    \item \textbf{Pubsub Systems}: 
    The top ten terms in this topic such as \textit{subscript}, \textit{subscribe}, and \textit{pubsub} suggest that posts are related to the GraphQL publish-subscribe paradigm (i.e., pubsub). GraphQL subscription is the third operation type that GraphQL supports, in addition to queries and mutations. Therefore, we map it to the (GraphQL) layer. In contrast with request-response models, GraphQL subscriptions follow a publish-subscribe mechanism where clients subscribe to the server and establish long-lived connections. On the other hand, the server is responsible for managing connections, routing, and publishing event messages to its subscribed clients. Our second-level LDA experiments produce four subtopics: Apollo client, WebSockets, Amazon Appsync, and event modeling. We find that the main protocol implementation of GraphQL subscriptions is WebSockets~\cite{websockets}, which ranks 7\textsuperscript{th} in the top-10 terms and appears in 15\% of the posts. Since WebSockets is a transport protocol, we map it to the (API Integration) component of our derived reference architecture. Amazon Appsync also appears in the top-10 terms and accounts for 34\% of the discussions. This finding suggests Appsync is the dominant backend for GraphQL subscriptions, whereas Apollo is the main client. Moreover, we observe that developers discuss libraries that support subscriptions (Framework). An example question from this topic asks \post{"I am using Appsync subscription for pub/sub messages. I have read some articles about subscriptions (like https://docs.aws.amazon.com/appsync/latest/devguide/real-time-data.html), and all of them mentioned specifying a GraphQL schema directive on a mutation. In my case, I have a lambda which is triggered by dynamodb streams and that lambda needs to publish the event to subscribers. How can I make this work in Appsync? I can't find any API I can use to publish event to Appsync"} [\hyperlink{https://stackoverflow.com/questions/62145701/how-can-i-publish-event-for-subscribers-in-appsync-graphql}{$Q_{62145701}$}].
    
    \item \textbf{Security Management}:
    This topic contains developer discussions about the security management of GraphQL on both the client and the server sides. This includes passing user ID and password, acquiring and managing JSON web tokens, passing authentication headers to the server, and validating user sessions on the server. The term "token" appears in 41\% of the posts. Our second-level LDA subtopics are \textit{User authentication}, \textit{User payload}, and \textit{User model}. The \textit{User authentication} subtopic discusses acquiring valid authentication tokens on the GraphQL client and passing them to the GraphQL server. The top terms in this subtopic include "authent", "compon", "link", and "header". The dominant framework to authenticate clients is React, as it is present in 29\% of the parent topic's posts. Additionally, it is incumbent to note that the GraphQL specification, which is protocol-agnostic by design, does not address authentication and rather leaves it as an implementation detail. Hence, GraphQL libraries such as Apollo-Link support authentication as a middleware adapted at the integration layer of the architecture. The \textit{User payload} subtopic contains user data passed between the GraphQL client and server. A few of the top 10 words of this subtopic include "firstname", "lastname", "regist", "phone" and "pass". Lastly, the \textit{User model} subtopic defines the server-side model used to resolve operations and fulfill authentication on user accounts. The top terms include "userid", "model", "account" and "session". Based on our analysis above, we map this topic to the \textit{Security layer} of our reference architecture. The following is an example post:  Another example is: \post{"I have a login which takes as inputs an email and password and when I press the button I have defined that the function handleLogin is called. The code of this function is as follows...If I type an incorrect username or password and press the button, I receive the corresponding "incorrect username or password" alert..."} [\hyperlink{https://stackoverflow.com/questions/74117267/problem-when-calling-function-in-react-repeated-result}{$Q_{74117267}$}].
    
    \item \textbf{E-commerce Product Management}:
    Posts on this topic concern the consumption of e-commerce management GraphQL APIs. Our second-level LDA experiments suggest two subtopics corresponding to \textit{Product Testing} and \textit{Product Selection}. 
    The \textit{Product Testing} subtopic discusses testing GraphQL services offered by popular e-commerce services such as Shopify to ensure that new listing and their updates are reflected accurately. 
    The \textit{Product Selection} subtopic discusses searching for e-commerce resources such as stores, products, tags, and variants using GraphQL queries.  Based on our analysis, we map this topic to the \textit{Client layer}.
    An example post is the following: \post{"I am working on fetching Shopify products based on the product tag. I did the below code for it, and it's working fine when I put the AND condition, But it does not work with the OR condition. It shows products where a tag appears in text content."} [\hyperlink{https://stackoverflow.com/questions/73063005/fetch-shopify-products-based-on-product-tags}{$Q_{73063005}$}].

    \item \textbf{GraphQL Client}:
    This topic contains developers' discussions regarding the use of the GraphQL client and various client libraries. The dominant GraphQL client is the Apollo client~\cite{apolloclient}, which we find present in 47\% of the posts. Apollo client represents the reference GraphQL client, which is implemented using the JavaScript React framework. This client features a set of components and techniques to allow local state management, data management, normalized in-memory caching, pagination, and more. It also provides various view integrations to frameworks such as Vue, Angular, and Ember. Our second-level LDA experiments produce four subtopics, namely: \textit{State management, Cache management, User interface}, and \textit{Apollo link libraries}. 
    \textit{State management} includes questions on local state management in GraphQL clients. 
    \textit{Cache management} ranges from simple operations, such as adding, updating, and removing data from the cache, to complex ones, such as defining cache policies, invalidation, optimistic responses, and normalization operations.  \textit{User interface} discusses data view management on the client side.  Apollo link libraries are fundamental to the architecture of the Apollo client. They help compose chains of functions to modify requests and responses as they pass through the client, such as HTTPLink, AuthLink, WSLink, and BatchLink. Based on our analyses of these four subtopics, we map this topic to the \textit{Client layer}. One SO example post is: \post{"...I would like the list to update and remove the deleted organization but that is not working...I have set up a cache exchange like this...I have (redundantly) tried two cache invalidation methods...The GraphQL query that returns... And the mutation to delete an organization looks like...cache.invalidate does not appear to do anything. I have checked the cache using the debugging plugin...I can see the records in the cache and they don't get removed."} [\hyperlink{https://stackoverflow.com/questions/72385225/urql-cache-invalidate-is-not-removing-items-from-the-cache}{$Q_{72385225}$}].

    \item \textbf{Prisma~\cite{prisma}}:
    This topic consists of posts regarding Prisma, an open-source server-side toolkit that instantly creates GraphQL APIs for existing databases using object-relational mapping (ORM). Prisma is implemented using NodeJS and Typescript. Since Prisma operates on the server side,  we map it to the \textit{Server layer} of the GraphQL architecture. Depending on the target database Prisma connects to, the toolkit automatically creates a specific Prisma schema using a Prisma modeling language. It also autogenerates a Prisma client to allow data to be retrieved from various database systems. Some databases it supports are MySQL, MongoDB, SQL Server, and PostgreSQL. GraphQL resolvers can use the Prisma client to expose the retrieved data as a GraphQL API. Prisma has two integration plugins to two GraphQL libraries (i.e., Nexus and TypeGraphQL) both of which help autogenerate GraphQL types and resolvers.  Our second-level LDA experiments produce two subtopics: \textit{Schema migration} and \textit{Deployment}. The first subtopic contains posts that concern database migration in Prisma using the command-line interface, i.e., Prisma CLI. The \textit{deployment} subtopic discusses deploying Docker images of Prisma servers to cloud platforms such as Heroku and Hasura~\cite{hasura}. The following is an example Prisma post on SO: \post{"what is the difference between prisma db push and prisma migrate dev ? When should I use one over the other. Docs say that prisma db push is about schema prototyping only and I don't understand what does that mean."} [\hyperlink{https://stackoverflow.com/questions/68539836/difference-between-prisma-db-push-and-prisma-migrate-dev}{$Q_{68539836}$}].

    \item \textbf{GraphQL Resolvers}:
    This topic contains questions and answers regarding fundamental server-side GraphQL resolver functions and schema type definitions, such as defining input and output types, fields, arguments, lists, queries, mutations, and resolver functions, in addition to managing and exporting schema files. Additionally, the topic involves discussions about user-defined types that GraphQL does not support natively (e.g., date types). We observe that developer discussions mostly concern server-side JavaScript applications predominantly utilizing the NodeJS framework and a few associated frameworks. This includes TypeScript, NestJS, Express, and Apollo Server. Moreover, developers discuss how to build GraphQL resolvers to fetch and manipulate data fetched from the database, and the most prevalent example is MongoDB using the Mongoose schema. The term "mongo" appears in 26\% of developer discussions within this topic. Therefore, we map it to the \textit{Server layer}. Our second-level LDA experiments confirm the above findings, presenting three subtopics, namely: \textit{TypeDefs}, \textit{NodeJS}, and \textit{Mongoose}. One example from SO is: \post{"I am creating a gql express server and i have created multiple schemas and referenced those schemas into each other. Assume collection USERS. Each user will have multiple challenges(schema or collection - challenge). I want to store the document(challenge) id into an array of challenges inside the user document that created it and vice-cersa. How can I acheive this? My user Schema...My challenge Schema...I have to also perform gql query to get the challenges details of the the user and also have to query all the users of a particular challenge"} [\hyperlink{https://stackoverflow.com/questions/69316367/i-have-multiple-mongoose-schemas-referenced-into-each-other-i-am-using-gql-comp}{$Q_{69316367}$}].
\end{enumerate}

\begin{boxedtext}
  \textbf{Takeaway 2.} Due to the complexity of the GraphQL ecosystem, developers discuss a diverse range of topics, including setup, migration, domain modeling, security, tooling, and API integration. 
\end{boxedtext}

\subsection{RQ3: How can we organize GraphQL discussions on StackOverflow
into a hierarchy of topics?}

\begin{figure}[ht]
  \centering
  \includegraphics[width=.8\textwidth]{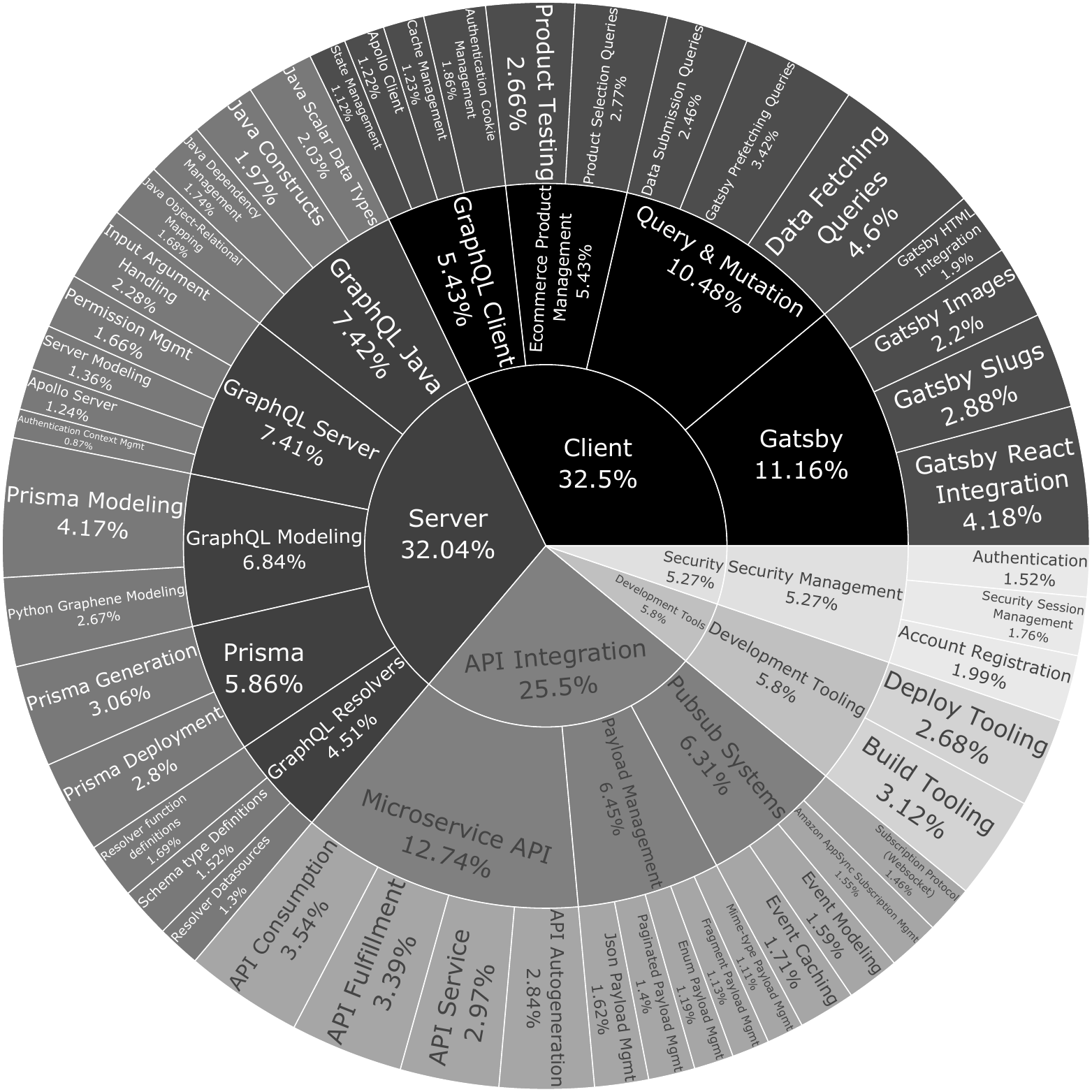}
  \caption{Hierarchy of GraphQL topics, their architecture layers, subtopics, and percentage of their questions.}
  \label{fig:topic-hierarchy}  
\end{figure}

Figure~\ref{fig:topic-hierarchy} is a sunburst chart that shows our GraphQL topic hierarchy consisting of three rings as follows. The innermost ring, which contains five architecture layers, corresponds to the GraphQL ecosystem and components we presented in Section~\ref{results_arch_layer}. The second ring consists of 14 segments corresponding to our discovered GraphQL topics listed in Table~\ref{table:topics}. The outermost ring contains 47 subtopics listed in Table~\ref{table:subtopics} where each set of subtopics belongs to a specific topic. Each topic segment has a parent architecture category and its own set of subtopics. For example, the topic \textit{GraphQL Java} belongs to the \textit{GraphQL server} category and has four subtopics, namely: \textit{scalar data types, constructs, dependency management} and \textit{object-relational mapping}. Furthermore, at any given level in the hierarchy, each segment is associated with a percentage representing its absolute size measured by its share of questions divided by the total number of questions. 

\begin{boxedtext}
  \textbf{Takeaway 3.} Software developers' discussions about GraphQL are grouped into a topic hierarchy of 5 architecture categories, 14 topics, and 47 subtopics. The 5 top-level categories are: \textit{GraphQL Client, GraphQL Server, API Integration, Security Management} and \textit{Development Tools}.
\end{boxedtext}

As shown, two-thirds of all GraphQL discussions concern either GraphQL server or client architecture layers, and each category segment amounts to almost one-third of the GraphQL discussions. On the other hand, GraphQL security and tooling are the least discussed GraphQL categories since either corresponds to about 5\% of all discussions. Finally, about 25\% of the discussions concern API integration between client and server.

\begin{boxedtext}
  \textbf{Takeaway 4.} About two-thirds of software developers' GraphQL discussions are equally contributed by GraphQL client and server categories.
\end{boxedtext}

 Even though the \textit{API Integration} layer ranks third in terms of its size among all architecture categories, we note that the \textit{Microservice API} topic represents the largest GraphQL topic, containing 12.74\% of the discussions. On the other hand, \textit{GraphQL Resolvers} is the least discussed topic at 4.5\% despite belonging to the most discussed GraphQL category, i.e., \textit{GraphQL Server}.

\begin{boxedtext}
  \textbf{Takeaway 5.} \textit{Microservice API} is the most discussed GraphQL topic, whereas the topic \textit{GraphQL Resolvers} is discussed the least.
\end{boxedtext}

\subsection{RQ4: How did GraphQL discussion topics on StackOverflow evolve?} 
\label{sec:popularity}

\begin{figure}[ht]
  \centering
  \includegraphics[width=\linewidth]{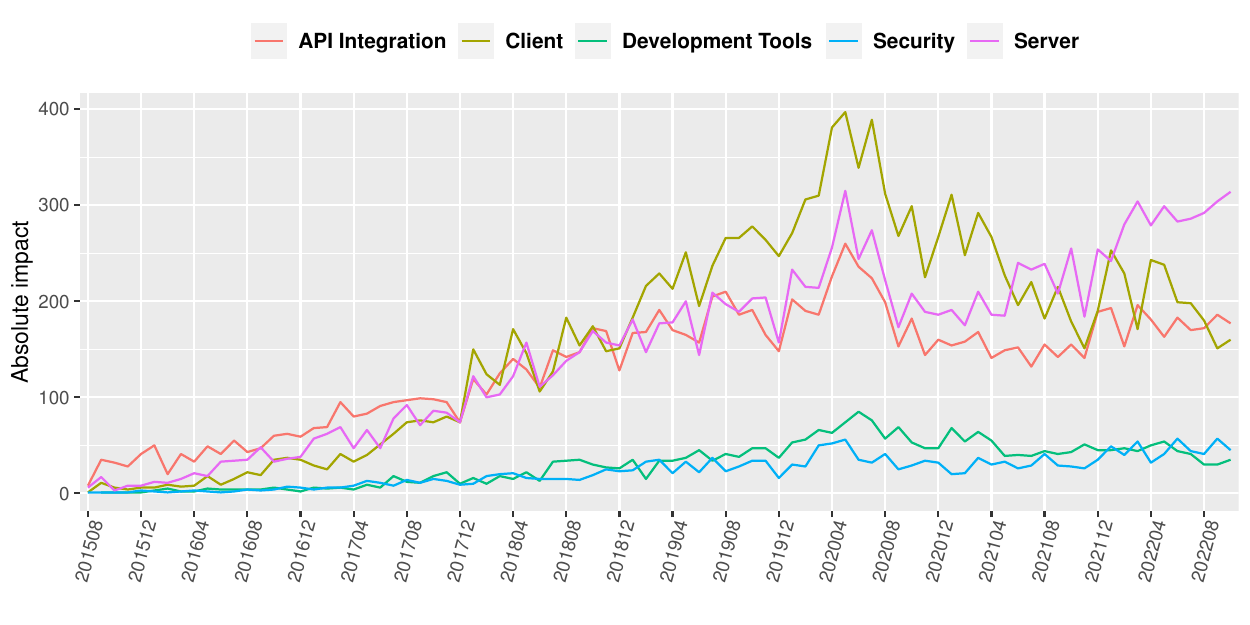}
  \caption{The absolute impact scores of GraphQL topics at various architectural layers in our dataset between August 2015 and November 2022.}
  \label{fig:absolute-impact}  
\end{figure}

To investigate the popularity of GraphQL, we analyzed the absolute and relative impact trends of GraphQL architecture categories. First, we calculated the absolute \textit{impact} (\ref{eq:absolute_impact}) of the GraphQL categories quarterly from August 2015 to November 2022, as shown in Figure~\ref{fig:absolute-impact}. We exclude December 2022 since that month's data is incomplete.
We observed an increase in the absolute impact for the categories \textit{GraphQL Client, GraphQL Server} and \textit{API Integration}. This increase is justifiable as the three categories correspond to new technologies proposed in the GraphQL specification. The increase in discussions may be attributed to several factors highlighting the growing interest, continued evolution, and adoption of GraphQL, especially as it peaked in December 2019 and August 2020. Some of these factors may include the announcement of the GraphQL Foundation and the establishment of the GraphQL specification project in 2019, the adoption of GraphQL by major companies, GraphQL conferences, and online events. The top contributor to the absolute impact during that period is the \textit{GraphQL Client}. Interestingly, the top contributor to the absolute impact changes starting around October 2021 to become the \textit{GraphQL Server}. This coincides with the release of the October 2021 GraphQL specification, which still represents the latest release when writing this paper.

On the other hand, the absolute \textit{impact} continued to stay stagnant for \textit{GraphQL Security} and \textit{Development Tools}. One explanation is that these categories represent areas of less GraphQL innovation. Concerning \textit{GraphQL Security}, the GraphQL specification claims a transport-agnostic approach that does not directly address security and leaves the details up to implementation protocols such as HTTP or WebSocket. Additionally, software developers' discussions under \textit{Development Tools} represent utilizing already existing tools (e.g., WebPack, NPM, and Yarn) to build and deploy GraphQL applications. 

\begin{boxedtext}
  \textbf{Takeaway 6.} According to absolute impacts, GraphQL client and server are the top two architectural layers attracting discussions on SO. Discussions of the GraphQL server are showing an increasing trend.
\end{boxedtext}

\begin{figure}[ht]
  \centering
  \includegraphics[width=\linewidth]{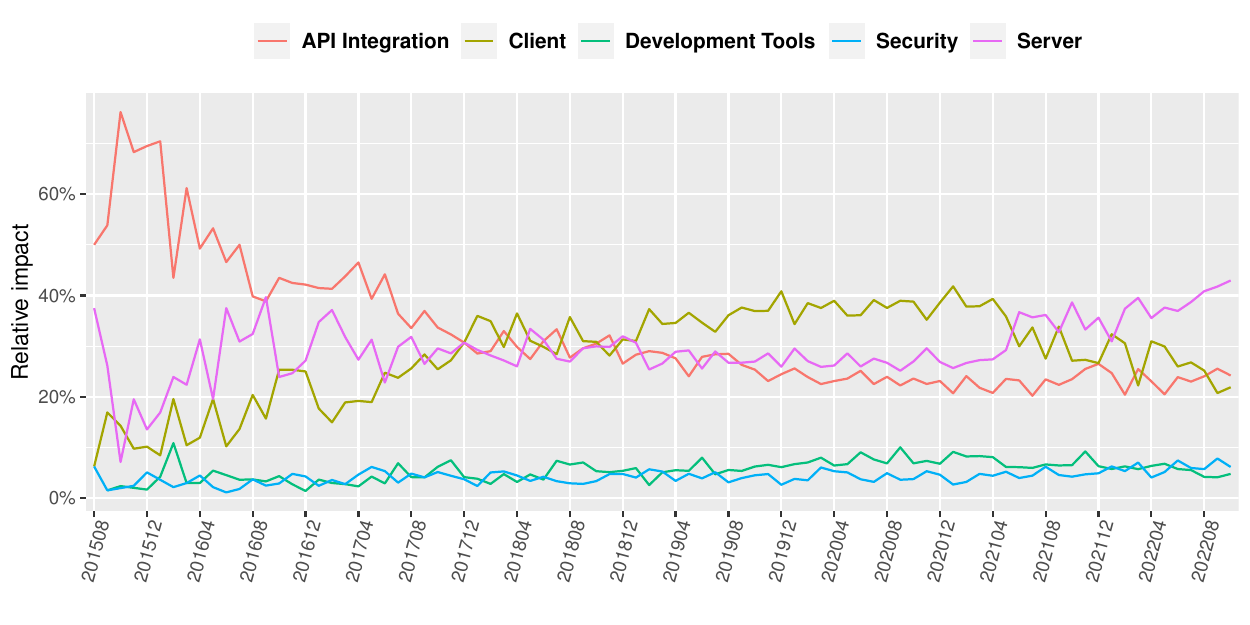}
  \caption{The relative impact scores of GraphQL topics at various architectural layers in our dataset between August 2015 and November 2022.}
  \label{fig:relative-impact}  
\end{figure}

We calculated the relative impact (\ref{eq:relative_impact}) of five GraphQL architectural layers quarterly from August 2015 to August 2022, as shown in Figure~\ref{fig:relative-impact}. Before December 2017, we observed a high relative impact of \textit{API Integration} that continues to decrease as the relative impact of both \textit{GraphQL Client} and \textit{GraphQL Server} increases. The initial intense change in the relative impact is likely associated with the public release of the GraphQL specification in 2015. In contrast, the continued changes in the relative impact until December 2017 are likely associated with the frequent releases and evolution of the GraphQL specification. The latter is an observation consistent at all points where the relative impact changes. We also observe a flat trend for both \textit{GraphQL Security} and \textit{Development Tooling}. This reinforces our initial explanation that these categories are areas of less GraphQL innovation and are, therefore, less impacted by GraphQL releases.

\begin{boxedtext}
  \textbf{Takeaway 7.} Discussions GraphQL API integration was majority till December 2017 for early adopters, who were integrating services offered by large organizations. In recent years, client and server-related discussions have seen increased shares, suggesting more organizations are implementing their services using GraphQL.
\end{boxedtext}

\subsection{RQ5: What are the difficulty and popularity of various GraphQL topics?} 
\label{sec:difficulty}
\begin{table}[ht]

    \centering
    \begin{tabular}{p{4.3cm}R{3cm}R{2cm} l}
    \hline
\textbf{Topic} & \textbf{\% W/O accepted answer} & \textbf{Delay hrs (median)} & \textbf{Layer} \\
\hline
Pubsub Systems &	0.69 &	18.14 & API Integration \\ 
GraphQL Client &	0.68	&13.22 & Client\\
Prisma &	0.68 &	10.16 & Server\\
Security Management &	0.65 &	4.01 & Security\\
Microservice API &	0.65 &	9.93 & API Integration\\
Development Tooling &	0.64 &	10.49 & Tooling\\
E-commerce management &	0.63 &	5.66 & Client\\
Java Implementation &	0.63 &	8.48 & Server\\
GraphQL Modeling &	0.63 &	10.32 & Server\\
Payload Management	& 0.62 &	6.33 & API Integration\\
GraphQL Server &	0.61 &	7.79 & Server\\
Query \& Mutation &	0.59 &	2.98 & Client\\
Gatsby&	0.57 &	5.10 & Client\\
GraphQL Resolvers &	0.56 &	3.56 & Server\\ \hline

\textbf{GraphQL (all)} &   \textbf{0.63} & \textbf{6.88} \\ \hline

    \end{tabular}
    \caption{Ranking of GraphQL discussion topics based on difficulty.}
    \label{tab:topic-difficulty}
\end{table}

Table~\ref{tab:topic-difficulty} ranks the GrapQL topics based on difficulty measured using \% of questions without accepted answers. We consider this metric a more accurate representation of topic difficulty since answer delay can be influenced by when a question is being asked (e.g., holiday, weekend, off-hours). Our results indicate that questions indicate that more than two-thirds of questions on \textit{Pubsub Systems}, \textit{GraphQL Client}, and \textit{Prisma} have no accepted answers. These three topics represent three different Architecture layers. Moreover, \textit{Security}, the fourth most difficult one, represents another layer. We also noticed that a ranking based on delay would also have \textit{Pubsub Systems} and \textit{GraphQL client} at the top two spots. 
On the other hand, \textit{GraphQL Resolvers}, \textit{Gatsby}, and \textit{Query \& Mutation} represent topics with the least difficulty both in terms of the ratio of questions with accepted answers and median answer delay.

\begin{boxedtext}
  \textbf{Takeaway 8.} \textit{Pubsub Systems}, \textit{GraphQL Client}, and \textit{Prisma} are three most difficult topics, while  \textit{GraphQL Resolvers}, \textit{Gatsby}, and \textit{Query \& Mutation} are three least difficult ones. 
\end{boxedtext}

\begin{table}[]

    \centering
    \begin{tabular}{p{4.3cm}R{2.5cm}R{2.5cm}l}
    \hline
       \textbf{Topic} &  \textbf{ View (median)}   &\textbf{Score (median)} & \textbf{Layer}  \\ \hline
GraphQL Client & 	458 &	1 & Client \\
Development Tooling &	440 &	1 & Tooling \\ 
Query  \& Mutation	&437	&1 & Client\\
GraphQL Resolvers &	433	&0 & Server\\
Payload Management &	423& 1 &  API Integration\\	
Java  Implementation&	416	&1 & Server\\
Pubsub Systems &	407	&1 & API Integration\\
GraphQL Modeling &	397	&1 & Server\\
Gatsby &	376 &	1 & Client\\
Security Management &	361 & 	0 & Security \\
GraphQL Server &	360	&1 & Server\\
E-commerce Management &	341 &	0 & Client\\
Microservice API &	328 &	1 & API Integration\\
Prisma &	323 &	0 & Server\\ \hline

\textbf{GraphQL (all)} & \textbf{387}& 1  \\ \hline
    \end{tabular}
    \caption{Ranking of GraphQL topics based on popularity}
    \label{tab:topic-popularity}
\end{table}

Table~\ref{tab:topic-popularity} ranks GraphQL topics based on median views.  \textit{GraphQL Client}, \textit{Development Tooling}, and \textit{Query \& Mutation} are the top three most popular topics. On the other hand, \textit{Prisma}, \textit{Microservice API}, and \textit{E-commerce management} are the three least popular topics. Our results do not indicate any association between popularity and Architecture layers, as the top five topics represent four layers.  
The median scores for all the topics are either 0 or 1, indicating no difference between the topics in terms of this measure. We want to mention that we do not use mean Scores, which have been used in many prior studies. We found that the Scores of the questions on SO are highly skewed, and few outliers can distort the means.

\begin{boxedtext}
  \textbf{Takeaway 9.} \textit{GraphQL Client} and \textit{Development Tooling} rank top in terms of median views, while \textit{Prisma} and \textit{Microservice API} rank bottom. 
\end{boxedtext}

\section{Discussion} \label{sec:discussion}
The following subsections detail key implications based on the results of this study.

\begin{figure}[ht]
  \centering
  \includegraphics[width=\linewidth]{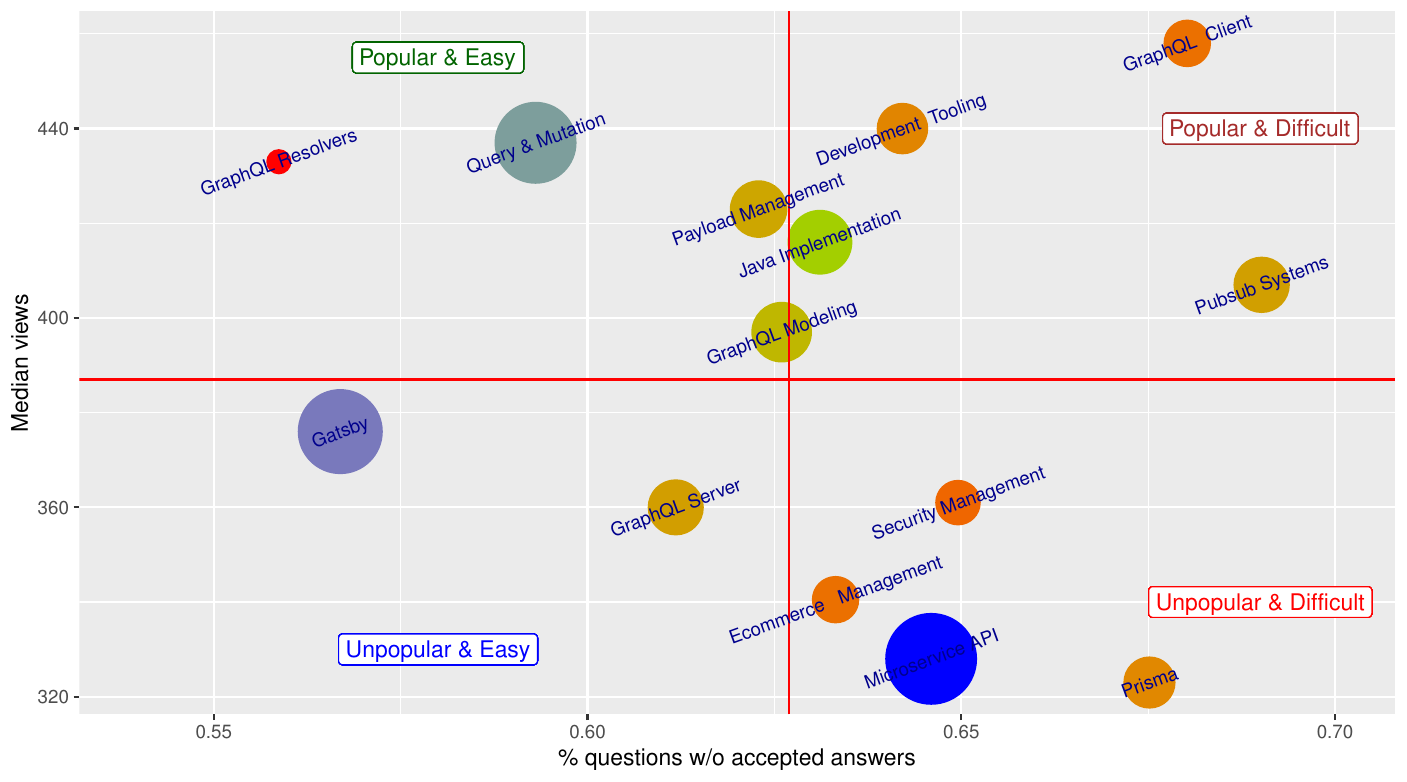}
  \caption{Significance of GraphQL topics using topic size and tradeoffs between topic popularity and difficulty.}
  \label{fig:tradeoff}  
\end{figure}

\subsection{Most Impacting Issues}
  Figue~\ref{fig:tradeoff} analyzes the tradeoff between the popularity and difficulty of the identified topics. The bubbles for each topic represent the topic size, defined as the percentage of SO questions contained within it. For example, the topic \textit{Microservice API} is the largest topic as it contains 12.7\% of the questions. In contrast, the topic \textit{GraphQL resolvers} is the smallest in size with a percentage of 4.5\%. The y-axis represents topic popularity measured by the median view count. The x-axis represents topic difficulty measured by the percentage of questions without an accepted answer. The plot also presents two median lines for all GraphQL questions (colored in red), dividing the plot into four quadrants, forming an effort-to-impact matrix~\cite{effortimpactmatrix}. We describe the quadrants as follows. 
  \begin{itemize}
      \item The \textbf{Popular and Easy} quadrant includes topics requiring less difficulty since they are more likely to have accepted answers. Yet, those are important for the GraphQL community due to the high number of views they receive. Four topics belong to this quadrant, which are \textit{GraphQL Resolvers}, \textit{Query \& Mutation}, \textit{Payload Management}, and \textit{GraphQL modeling}. Therefore, GraphQL experts who want to improve their reputation score of SO can focus on these topics to maximize their return on investment (ROI).

\item  The \textbf{Popular and Difficult} quadrant represents difficult topics due to the lower probability of having accepted answers. However, these topics are significant due to high community interests. Hence, GraphQL framework builders should prioritize improving support and documentation on these four topics, which include \textit{GraphQL Client}, \textit{Pubsub Systems}, \textit{Development Tooling}, and \textit{Java Implementation}. Although these topics represent four different architecture layers, they relate to implementing new applications using GraphQL.

\item The \textbf{Unpopular and Easy} quadrant contains GraphQL topics that are more likely to be answered and have a lower community impact than the topics from the popular half. Two topics belong to this quadrant, which are \textit{Gatsby} and \textit{GraphQL server}. According to our results, these topics do not require immediate attention. 

\item Finally, the \textbf{Unpopular and Difficult} quadrant contains difficult topics with less community impact. Four topics belong to this quadrant: \textit{Security management}, \textit{Prisma}, \textit{E-commerce management}, and \textit{Microservice API}.  Although these topics currently have a lower community impact, the number of questions on these areas is significant. Noticeably, Microservice API, which has the highest ratios of questions, belongs here. Hence, although many users are not encountering challenges in these areas, many unresolved issues remain. Since GraphQL adoption is a rapidly growing technology, these challenges will be encountered by more community members. Hence, topics from this quadrant should receive the second most priority from GraphQL framework builders. 
      
  \end{itemize}

   \subsection{Implications for Practitioners} 
   Our analysis of popularity (Table~\ref{tab:topic-popularity}) and difficulty (Table~\ref{tab:topic-difficulty}) suggest that tools from the GraphQL ecosystem lack adequate support as approximately two-thirds of questions on \textit{Prisma} and \textit{Development Tooling} do not have accepted answers. Our analysis suggests that many questions on these topics inquire how to achieve particular objectives using these tools. For example, \post{``How to use mongodb's increments operator (\$inc) in prisma ORM? I cannot find a way to use some special MongoDB operation on prisma: In MongoDB I can do this:"}[\hyperlink{https://stackoverflow.com/questions/73556355/how-to-use-mongodbs-increments-operator-inc-in-prisma-orm}{$Q_{73556355}$}]. Automated tools play an important part in promoting the adoption and ease of onboarding of new practitioners.  Hence, GraphQL tool builders should focus on creating FAQs and examples detailing such common usage patterns.

 As discussed in Section~\ref{sec:related_work}, a GraphQL server supports three operations, i.e., Query, Mutations, and Subscription. Our results indicate that while questions on \textit{Queries \& Mutations} fall under the lower end of difficulty spectra, \textit{PubSub Systems} is one of the most difficult topics. Despite having thousands of views, indicating popularity, the majority of questions on subscription do not have accepted answers. Our analysis suggests that many questions are about setting up and testing basic subscriptions. For example, 
 \post{``I have a secure subscription endpoint on my Apollo server. I can send subscription connectionParams from my react client by setting it in the WebSocketLink constructor and verify it in the onConnect property of Apollo Server constructor. But how do I test these subscriptions on GraphQL Playground?"}[\hyperlink{https://stackoverflow.com/questions/63635513/graphql-playground-how-to-set-subscription-connectionparams}{$Q_{63635513}$}]. Hence, we recommend GraphQL foundation to focus on improving documentation, support, and tutorials on subscription operations. 

For managers considering an adoption, GraphQL specification is language agnostic, and GraphQL has implementations in popular programming languages such as  Node.js, Java, Python, Go, .Net, and Rust. However, our results suggest that not all implementations have the same level of support and popularity. Since the reference implementation uses JavaScript (i.e., Node.js), this ecosystem has matured tools and better community support. While an enterprise-based application may prefer Java due to performance and existing developers' familiarity with the Java ecosystem, our results indicate that Java implementation currently has less support than JavaScript-based ones. 
Moreover, GrahpQL client-related areas such as \textit{GraphQL Resolvers} and \textit{Query and Mutation} are less difficult, yet popular. Based on this result, we recommend newcomers onboarding to the GraphQL ecosystem begin with these areas to understand basic building blocks and usage of GraphQL before moving to a more challenging topic such as \textit{GraphQL server} and \textit{GraphQL Modeling}.

    \subsection{Implications for Researchers} 
   The design of GraphQL imposes unique security challenges~\cite{mcfadden2024wendigo,yazdipour2020github,wittern-2019}. For example,  GraphQL allows for query batching, where multiple queries are sent in a single request, and aggregation, where multiple pieces of data are fetched together. These features can be exploited to perform more efficient attacks, such as extracting large volumes of data in a single request, leading to Denial of Service attacks.
   Moreover, implementing fine-grained access control in GraphQL can be challenging because of its flexible nature. Ensuring that users can only access the data they are authorized to see requires careful design and enforcement of authorization rules at the field level.
   Our results indicate that i) only 5\% questions are on security, ii) Security is not a popular topic in terms of views, and iii) security remains the fourth most difficult topic in terms of questions without accepted answers. These results may indicate that developers often do not have adequate knowledge or even no solutions to implement security in GraphQL server implementations. In some cases, although developers want to implement security, existing frameworks lack support. For example, \post{``I have a very straight forward graphQL app built on top of AWS AppSync, and I want to configure support for HTTP Strict Transport Security (HSTS) so that's my question how can I enforce a support HTTP Strict Transport Security (HSTS) on AWS AppSync or how can I add an HTTP header in the AppSync response"} [\hyperlink{https://stackoverflow.com/questions/63347634/how-to-enforce-http-strict-transport-security-hsts-on-aws-appsync-graphql-resp}{$Q_{63347634}$}]. Since is expected to see rapid adoption across various industries~\cite{ibm_gartner_blog_post}, we recommend researchers to explore solutions to integrate security at various GraphQL layers.

\section{Threats to Validity} \label{sec:threats}
\vspace{2pt}
\noindent \textbf{Internal validity}:
Naming our discovered topics is another threat that may impact the validity of our study, especially since we performed our naming process manually. To minimize such threat we used independent naming by three different authors, each with extensive experience in GraphQL development. Conflicts were resolved through discussion. Additionally, we conducted second-level LDA experiments to identify subtopics within each topic category. The subtopics help produce more insight into the contents of the topics rather than solely relying on the top words within their respective parent topics. We also applied well-known techniques~\cite{bigdata, concurrency} to validate the topic names such as selecting random questions and answers from SO.

\vspace{2pt}
\noindent \textbf{Construct validity}:
We consider our decision to utilize LDA to perform topic modeling a threat since LDA is prone to stability and coherence issues~\cite{ldade}. To mitigate that we apply standard techniques~\cite{bigdata, concurrency, web, mobile} in conjunction with exhaustive experiments. For example, we repeat every single LDA experiment 10 times in order to capture the results with the highest coherence values. We also ensure the stability of our repetitions by utilizing the Jaccard similarity score. 
  Deciding to include both the titles and bodies of the SO questions and answers as opposed to only the titles poses another threat. Therefore, we employed well-known preprocessing steps to cleanse the posts and help reduce the noise before qualifying the documents into our LDA system. 
 Yet another threat is determining the optimal number of topics K for both top-level and second-level LDA experiments, especially since landing an optimal K value is cumbersome~\cite{bigdata}. To minimize that, we exhausted all possible K values in a brute-force fashion, stepping the number of topics by 1 each time, at the expense of experimentation time and effort. Deriving a proper GraphQL reference architecture is also a threat which, in order to alleviate, we applied standard reference architecture derivation techniques~\cite{web_browser_ref_arch, web_server_ref_arch, blockchain} and also utilized two concrete GraphQL implementations. Sound mapping of topics to the respective architecture layers of the GraphQL ecosystem is a threat due to variations in topic sizes and mapping cardinality. To reduce such a gap, we conducted second-level LDA experiments and performed architecture mapping based on additional insights produced from the subtopics.

\vspace{2pt}
\noindent \textbf{External validity}: Using SO as the only source of our dataset is a threat that may compromise the external validity of our results. However, SO is the most popular and goto place for software developers to find solutions to technical challenges. Moreover, 
our dataset is large with approximately 45k of real-world questions and answers to provide crucial insights. 

\vspace{2pt}
\noindent \textbf{Conclusion validity}: We conducted Anderson–Darling tests~\cite{anderson1952asymptotic} to test normality of popularity and difficulty measures. Since our results suggest non-normal distributions, we refrain from reporting potentially misleading mean measures. Hence, we do not anticipate any threat arising from our statistical tests. 

\section{Conclusion} \label{sec:conclusion}

GraphQL is an emerging technology, aiming to serve as an alternative or coexisting API to legacy protocols that dominated the industry for the past decade (e.g., REST). Due to its flexibility and efficiency, it has been gaining rapid adoption among various industries~\cite {ibm_gartner_blog_post}. With the goal of understanding challenges encountered by GraphQL developers, we analyzed 45K StackOverflow questions and answers on GraphQL.  We conducted a mixed-method empirical analysis, where we leveraged LDA to automatically identify 14 topics and 47 subtopics. We derive a reference architecture for the GraphQL ecosystem with five key layers based on two well-known GraphQL implementations. Furthermore, we manually inspect questions on each topic and subtopics to create a topic-to-architecture mapping and provide additional insight to the GraphQL stakeholders.  Our results indicate that \textit{Client} and \textit{Server} are the top two architectural layers attracting discussion on SO. While earlier discussions on SO focused on building third-party applications consuming GraphQL APIs (i.e., API Integration) released by large organizations, recent trends suggest more organizations implementing APIs using GraphQL servers. Discussions on basic building blocks of GraphQL, such as \textit{Resolvers}, \textit{Query \& Mutations}, \textit{GraphQL modeling}, and \textit{Payload Management} are both popular and are better supported by the community. On the other hand, implementations of new GraphQL servers, clients, managing subscriptions, and tooling have a lower level of support, although those areas have high community interests.  Due to difficulty and lack of well-defined solutions, security remains difficult and low-interest area, which can lead to vulnerabilities~\cite{mcfadden2024wendigo,yazdipour2020github}. Hence, efficient implementation of Security at various layers is a pressing research need. 

\section*{ Data availability}\label{Data_availability}
{ We have made our dataset, source code, and results publicly available on GitHub at: \url{https://github.com/WSU-SEAL/graphql-topic-modeling}}

\bibliographystyle{ACM-Reference-Format}
\bibliography{references}
\end{document}